\documentclass[prb,aps,twocolumn,showpacs]{revtex4}
\usepackage{amsmath,amssymb,amsfonts,float}
\usepackage[dvips]{graphicx,color}    
\usepackage{bm}
\usepackage{times}       
\usepackage{ulem}
\usepackage{pifont}
 
\newcommand{\rmS}{\text{\rm S}}

\newcommand{\bolk}{\mathbf{k}}
\newcommand{\bolq}{\mathbf{q}}

\newcommand{\bolQ}{\mathbf{Q}}

\newcommand{\boll}{\mathbf{l}}
\newcommand{\bolm}{\mathbf{m}}

\newcommand{\ket}[1]{| #1 \rangle}  
\newcommand{\VEV}[1]{\langle #1 \rangle}  

\newsavebox{\dotdot}
\savebox{\dotdot}[3mm]{\shortstack{\circle*{0.8}\\ \\ \circle*{0.8}}}

\begin{document}
\title{
Supersolid phase of Three-dimensional spin- and hardcore-boson models}
\author{Hiroaki~T.~Ueda and Keisuke~Totsuka}
\affiliation{Yukawa Institute for Theoretical Physics, 
Kyoto University, Kitashirakawa Oiwake-Cho, Kyoto 606-8502, Japan}
\begin{abstract}
We study the stability 
of solid- and supersolid (SS) phases of a three-dimensional 
spin- and a hardcore-Bose-Hubbard models 
on a body-centered cubic lattice. 
To see the quantum effects on 
the stability of the SS phase, we model the vacancies 
(interstitials) introduced in the solid, which are believed responsible for 
the appearance of the SS phase, by spinwave bosons and adopt 
the interaction 
between the condensed bosons as a criterion. 
A repulsive nature of the low-energy effective interaction 
is the necessary condition 
for a second-order solid-SS transition and, 
when this condition is met, normally the SS phase is
 expected. 
In calculating the effective interaction, 
we use expansions from the semiclassical- (i.e. large-$S$) 
and the Ising limit 
combined with the ladder approximation. 
The impact of quantum fluctuations crucially depends on 
the energy of the solid phase and that of the superfluid phase 
at half filling. 
As an application to $^4$He, we study the parameter region 
in the vicinity of the fitting parameter set given by Liu and Fisher. 
For this parameters set, quantum fluctuations 
at the second order in $S^{-1}$ destabilize 
the solid phase, which is supposed to be stable within the mean field theory. 
\end{abstract}            
\maketitle
\section{Introduction}
\label{sec:intro}
The supersolid (SS) state, which has both diagonal- and 
off-diagonal long-range order, has been investigated over the past 
five decades\cite{Penrose,Andreev,Chester}.  
Recently, Kim and Chan 
suggested\cite{Kim-Chan} that the observed non-classical rotational inertia 
(NCRI)\cite{Leggett} in solid $^4$He, might be attributed  
to coexisting superfluidity. 
This experiment sparked a renewed interest 
and the origin of the NCRI is still under debate\cite{debate}.

The quantum lattice gas model (QGM)\cite{Matsubara-Matsuda}, 
or equivalently the hardcore-Bose-Hubbard model, is one of the simplest  
models suited for studying the low-temperature physics of quantum solids. 
Since the QGM is in an exact correspondence 
to $S=1/2$ quantum spin models\cite{Matsubara-Matsuda}, 
we can use powerful methods 
developed in quantum spin systems in understanding the physics underlying 
the QGM. 
The QGM has been applied\cite{Matsuda-Tsuneto} 
to study the possibility of the SS in $^4$He, 
and later the comprehensive discussion\cite{Liu-Fisher} given 
by Liu and Fisher concluded, within the mean-field approximation (MFT), 
that the SS exists in $^4$He. 
However, recent studies on the SS in the 2D square 
lattice systems revealed that quantum fluctuations 
dramatically change the behavior and may even suppress 
the SS which is supposed to exist 
within the MFT\cite{Batrouni-3,Batrouni-2,Dang}. 
For the optimal fitting parameter set 
obtained by Liu and Fisher for $^4$He ({\it LF point}; 
see (\ref{LiuFisherSS})), frustration seems to play an important role. 
Hence, interplay between quantum fluctuations 
and frustration may change the physics of the QGM of $^4$He. 

Recently, Bose-Einstein condensation (BEC) 
of magnons has been observed 
experimentally\cite{Nikuni-O-O-T-00,RaduMagBEC} and 
is now widely investigated\cite{Giamarchi-R-T-08}. 
Effects of frustration on magnon BEC 
would be intriguing in their own right, 
as frustration may enhance quantum effects and even lead to 
such exotic condensed states 
as the SS which are hardly realized in real Bose systems. 
For instance, quite recently, Takigawa {\it et al.} 
reported\cite{SrCu2(Bo3)2-SS-2} 
a persisting spin superlattice 
in SrCu$_2$(BO$_3$)$_2$ (SCBO)\cite{SrCu2(Bo3)2} 
coexisting with (possibly) mobile magnons even beyond 
the $1/8$-plateau, which is
reminiscent of the SS state 
predicted theoretically\cite{SrCu2(Bo3)2-SS-1} for SCBO. 
Because weak anisotropic interactions break the rotational symmetry 
around the externally-applied magnetic field (or, U(1) gauge 
symmetry in the QGM language), this phase 
may not a true SS phase. 
However, the discrete subgroup of the rotational symmetry 
can be spontaneously broken\cite{Penc-07} and the observed phase might 
still hold a close relationship to the SS in its original sense. 
The physics of this phase and the realization of the `magnon SS' in
other 
compounds are also topics to be investigated more closely. 

For the clear understanding of NCRI in $^4$He 
and the SS states 
in spin systems, it is useful to find a criterion which 
assesses the combined effect of quantum 
fluctuations 
and frustration on the stability of the SS phase. 
In this paper, with the help of spin
wave expansion, we push ahead with 
the widely accepted intuitive picture\cite{Andreev,Chester} 
that BEC of vacancies or interstitials gives rise to the SS state to 
propose that the interaction among the 
condensed vacancies (interstitials) serves as a good criterion 
for the stability of the SS. 
To this end, we adopt the so-called dilute-Bose-gas 
technique\cite{Beliaev-58}. 
Normally, the dilute-Bose-gas approach is used 
{\it only} in the vicinity of the saturation field 
to obtain unbiased (asymptotically) exact 
results\cite{Batyev,Nikuni-Shiba-2,VeilletteandChalker,Chubukov,HTUandKT},  
since the lack of an exact reference state (i.e. vacuum) on which 
boson excitations are defined hampers the construction of a well-defined 
bosonic Hamiltonian. 
To overcome this difficulty, we introduce the spin magnitude $S$ 
and the Ising-like anisotropy as large control parameters, which 
guarantee the validity of the reference state 
even far below the saturation field, 
and develop a systematic expansion with respect to these parameters.   

\begin{figure}[H]
\begin{center}
\includegraphics[scale=0.4]{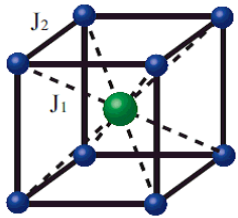}
\caption{(color online) 
Three-dimensional body-centered cubic (bcc) 
lattice and interactions considered in the text. 
Filled circles denote spins connected by anisotropic 
(XXZ-like) exchange interactions. 
We divide the lattice into two sublattices, which are 
distinguished by the size of spheres.  
Each sublattice forms a simple cubic lattice.
\label{Fig;bcclattice}} 
\end{center}
\end{figure}

The organization of the present paper is as follows. 
In Sec.~\ref{Sec;model}, we introduce a 
three-dimensional model Hamiltonian on a body-centered cubic
(bcc) lattice (see Fig.~\ref{Fig;bcclattice}) and briefly review the
correspondence between the spin model and the QGM.  
At the same time, we classify the ground-state phases within the MFT.  
Then, we derive a spin-wave Hamiltonian by using the
Dyson-Maleev transformation in the solid phase. 

In Sec.~\ref{Sec;general}, we outline the dilute-Bose gas approach 
used in investigating the SS phase around the solid phase. 
If the effective interaction among the condensed bosons 
is attractive, 
the SS phase for low condensate density is normally 
phase-separated. 
Although we do not exclude the possibility that the SS emerges 
through the first-order transition from the solid phase, 
this seems unlikely from various results obtained 
by quantum Monte-Carlo 
simulations\cite{Batrouni-3,Batrouni-2,suzuki,Mila,dimerMC,Dang}. 
Hence, in this paper, the SS is said to be `unstable' (`stable') 
if the interaction between condensed bosons is attractive (repulsive). 
To evaluate the interaction concretely, 
we need approximations. 
In Sec.~\ref{LargeS}, we study the properties of the solid and 
the SS phase by the large-$S$ expansion up to the second order 
in $S^{-1}$. 
At the first order, the MFT results are 
reproduced.
We shall find three types of SSs, 
which have properties 
similar to those appearing in the 2D-square 
lattice\cite{Batrouni-4,Batrouni-1,Mila,Pich-Frey}.  
The formulation of the second-order perturbation 
is detailed. 

In Sec.~\ref{LargeJ}, we study the properties of the solid- 
and the SS phase by the Ising expansion up to the second order. 
Although quantum fluctuations seem to suppress the interactions 
at the first order, 
the boundary determining the stability of the SS does not shift 
and the magnetization process is affected only quantitatively 
by quantum fluctuations. 
In other words, 
the stability itself is known from the MFT 
if the large Ising anisotropy exists.  
To see the effect of quantum fluctuations on the stability of SS, 
we have to proceed to the second order calculation. 

Our main results are summarized in Sec.~\ref{sec:lattice}, 
where we study the stability of the solid and 
the SS phases focusing on the LF point. 
Readers who only want to know the main results may skip 
Sec.~\ref{LargeS} and \ref{LargeJ} and go directly to 
this section. 
For the parameter set corresponding to the LF point, 
quantum fluctuations destabilize 
the solid state expected from the MFT 
(Fig.~\ref{Fig;spinwaveGround-2}) at least within the 
conventional second-order spin-wave expansion. 
Concerning the stability of the SS, 
both of the two second-order calculations 
conclude that quantum fluctuations only slightly change 
the MFT boundary of the SS phase, provided that 
the energy of the solid phase is sufficiently smaller than that of 
the superfluid phase at half filling (Fig.~\ref{Fig;GammaS2I2}).  
In the vicinity of the LF point, where the above condition 
is not satisfied, 
it is suggested that the SS phase is fragile against 
quantum corrections or even completely smeared out, 
although the validity of both approaches is not obvious in this region. 

For concreteness, we restrict our discussion in this paper 
to a quantum spin model on a bcc lattice.  
However, our approach can be easily generalized to 
quantum spin models on other 3D lattices. 

\section{Spin Hamiltonian and Quantum lattice gas model}
\label{Sec;model}
\subsection{Model Hamiltonian}
\begin{figure}[H]
\begin{center}
\includegraphics[scale=0.4]{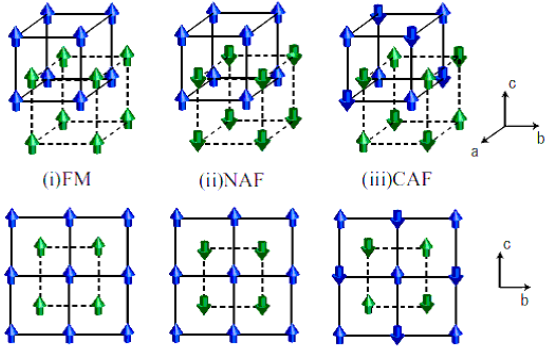}
\caption{(color online) 
Spin configurations for the three phases (`FM', `NAF' and `CAF') 
in the text. 
(i) `FM' (ferromagnetic phase) represents a phase where all spins 
are polarized along the field direction. 
(ii) In `NAF', the spins on each sublattice align 
ferromagnetically while those on different sublattices are anti-parallel. 
(iii) `CAF' is made up of two antiferromagnetically-ordered sublattices, 
which, as a whole, align in a collinear manner. 
\label{Fig;config}} 
\end{center}
\end{figure}

Let us consider the following frustrated spin Hamiltonian on the bcc lattice with the nearest neighbor Ising antiferromagnetic (AF) interactions ($J_1^z>0$):
\begin{equation}
\begin{split}
H&=\sum_{\text{n.n.}}\left\{ J_1^z S_{\bf i}^zS_{\bf j}^z
+J_1^\perp(S^x_{\bf i}S^x_{\bf j}+S^y_{\bf i}S^y_{\bf j})\right\} \\
&+\sum_{\text{n.n.n.}}\left\{ J_2^z 
S_{{\bf i}^\prime}^zS_{{\bf j}^\prime}^z
+J_2^\perp (S^x_{{\bf i}^\prime}S^x_{{\bf j}^\prime}
+S^y_{{\bf i}^\prime}S^y_{{\bf j}^\prime})\right\}\\
&+Sh\sum_{{\bf i}} S_{{\bf i}}^z\ ,
\end{split}
\label{HSpin}
\end{equation}
where the summations n.n. and n.n.n. are taken for the
nearest-neighbor and the second-nearest-neighbor pairs, respectively. 
This Hamiltonian in the case that $J_i^z=J_i^\perp$ for $i=\{1,2\}$ 
(Heisenberg case)
has been investigated from the various approaches.\cite{bccHeisenberg} 
In the case of $S=1/2$, this Hamiltonian is equivalent to the following
hard-core bosonic Hubbard model,\cite{Fisher-Hubbard} 
\begin{equation}
\begin{split}
H&=\sum_{\text{n.n.}}\left\{\frac{J_1^\perp}{2}(p^\dagger_{\bf i} p_{\bf j}+p_{\bf i} p_{\bf j}^\dagger)+J_1^z \hat{n}_{\bf i}\hat{n}_{\bf j}\right\} \\
&+\sum_{\text{n.n.n.}}\left\{\frac{J_2^\perp}{2}(p^\dagger_{{\bf i}^\prime} p_{{\bf j}^\prime}+p_{{\bf i}^\prime} p_{{\bf j}^\prime}^\dagger)+J_2^z \hat{n}_{{\bf i}^\prime}\hat{n}_{{\bf j}^\prime}\right\}\\
&-\mu_{h}\sum_{\bf i} \hat{n_{\bf i}}\ ,
\end{split}
\label{eqn:Bose-Hubbard}
\end{equation}
where $\hat{n_{\bf i}}=p^\dagger_{\bf i} p_{\bf i}$. This model can be 
used to study the low-energy physics of $^4$He if we approximate the Bose gas by
the QGM\cite{Matsubara-Matsuda}. 
Specifically, the `longitudinal' couplings $J_{1,2}^z$ and `transverse'
ones $J_{1,2}^\perp$ mimic the interaction potentials and the kinetic energy
of Helium, respectively, and the external magnetic field $h$
(or $\mu_h$) controls the pressure. In the QGM, $J_{1,2}^\perp < 0$ and $J_1^\perp/J_2^\perp $ is fixed at $1/2$ because of the lattice structure.
Liu and Fisher suggested several sets of fitting parameters appropriate
for $^4$He and concluded that the stability of the SS phase is ensured
within the MFT\cite{Liu-Fisher}. However, the existence of quantum
fluctuations and frustration effects may destroy the classical
ground state. To see the validity of the MFT, in Sec.~\ref{sec:lattice}
we shall study these effects on the ground state in the vicinity of the
following parameter set (LF point; the case (a) 
in Ref.~\onlinecite{Liu-Fisher}):
\begin{equation}
J_1^z=2.60,\ J_2^z=1.59,\ J_1^\perp =-1,\ J_2^\perp =-0.5\ .\label{LiuFisherSS}
\end{equation}
Since the Ising-like N\'{e}el antiferromagnetic (NAF) phase is 
identified with a solid phase of $^4$He, 
we restrict ourselves only to the case 
that NAF order along the $z$-direction appears around $h= 0$ 
and will not consider the Ising-like collinear antiferromagnetic (CAF)
phase which realizes, in the classical case, when $2J_1^z<3J_2^z$.  
Let us briefly discuss possible classical phases at $h=0$.  
The classical phases fall into three fundamental classes (NAF, CAF, FM) 
as is shown in Fig.~\ref{Fig;config}.  
These phases are further classified by whether the spins 
align along the $z$-axis or in the $xy$-plane. 
In the former case, the ground state may be gapped.
In the latter case, the spontaneous symmetry breaking of the rotational 
symmetry around the $z$-axis ($U(1)$) induces 
the gapless Goldstone mode and the phase is viewed as a 
superfluid (SF). 
The $U(1)$-broken phases accompanied by 
translation-symmetry breaking in the {\it diagonal} channel 
(i.e. $\VEV{p^\dagger_\boll p_\boll}$ or $\VEV{S^z_\boll}$) 
as well are thought of as spin-analogues of SSs.\cite{Batrouni-4,Batrouni-1}

When the spins align in $z$-axis, the energy of each Ising-like phase is given by
\begin{subequations}
\begin{align}
\frac{E_{\rm Ising-NAF}}{2NS^2}&=-4J_1^z+3J_2^z\ ,\\
\frac{E_{\rm Ising-CAF}}{2NS^2}&=-3J_2^z\ ,\\
\frac{E_{\rm Ising-FM}}{2NS^2}&=4J_1^z+3J_2^z-|h|\ ,
\end{align}
\label{EIsing}
\end{subequations}
where $N$ is the number of sites of each sublattice. 
When the spins align in $xy$-plane, the phases are viewed as 
SFs and the energy of each phase is given by
\begin{subequations}
\begin{align}
\frac{E_{\rm xy-NAF}}{2NS^2}&=-4J_1^\perp+3J_2^\perp\ ,\\
\frac{E_{\rm xy-CAF}}{2NS^2}&=-3J_2^\perp\ ,\\
\frac{E_{\rm xy-FM}}{2NS^2}&=4J_1^\perp+3J_2^\perp\ .
\end{align}
\label{ESF}
\end{subequations}

\begin{figure}[H]
\begin{center}
\includegraphics[scale=0.43]{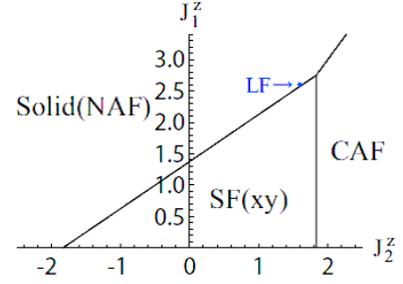}
\caption{(color online) Classical phase diagram for $h=0,\ J_1^\perp =-1,\ J_2^\perp =-1/2$. NAF and CAF are implied as the Ising-like gaped ones. The dot labeled as LF represents the LF point (\ref{LiuFisherSS}).
\label{Fig;QGMCl}} 
\end{center}
\end{figure}

The ground-state phase diagram of the QGM for $h=0$ and $J_1^\perp/J_2^\perp =1/2$ is shown in Fig.~\ref{Fig;QGMCl}.

\begin{figure}[H]
\begin{center}
\includegraphics[scale=1.0]{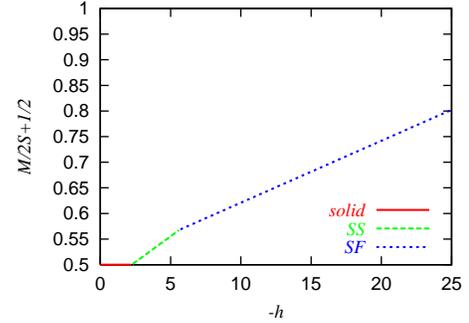}
\caption{(color online) The magnetization curve obtained 
for the LF point (\ref{LiuFisherSS}) within the MFT. 
Magnetization $M$ is given by $(1/2N)\sum_\boll^{2N} \VEV{S^z_\boll}$. 
The existence of the phases solid, SS and SF is confirmed.
\label{Fig;ClLF}} 
\end{center}
\end{figure}

The magnetization curve for the LF point (\ref{LiuFisherSS}) is shown in Fig.~\ref{Fig;ClLF}.

To see the physics of the NAF phase more clearly, we divide the whole lattice into two sublattices $A$ and $B$ each of which forms a cubic lattice. Since we can change the sign of the (n.n.) transverse coupling $J_1^\perp \rightarrow -J
_1^\perp$ at will by making spin rotation (by $\pi$) around the $z$-axis
$S_l^i \rightarrow -S_l^i(i=x,y,l\in A)$ only for the A-sublattice, 
we may restrict our consideration to the case $J_1^\perp \leq0$. 

The correspondence between the phases in the quantum
lattice-gas formulation and the ones in the (quantum) spin-model 
formulation is shown in TABLE \ref{table:correspondence}. 
\begin{table}
\caption{Correspondence between 
the quantum lattice-gas model (QGM) and the spin model. 
`TS' denotes translational symmetry.}
\label{table:correspondence}
\begin{center}
\begin{ruledtabular}
\begin{tabular}{cc}
\raisebox{0.5ex}[0pt]{QGM(Bose-Hubbard model)} & 
\raisebox{0.5ex}[0pt]{Spin model} \\ 
\hline
vacuum & (polarized) FM \\ 
checkerboard solid & (Ising-like) NAF \\ 
striped solid & (Ising-like) CAF \\ 
SF ($\VEV{p_{\bf i}p_{\bf j}}\neq 0$ with TS) &
$\VEV{S^{+}_{\bf i}S^{+}_{\bf j}}\neq 0$ with TS \\ 
SS ($\VEV{p_{\bf i}p_{\bf j}}\neq 0$ with broken TS) &
$\VEV{S^{+}_{\bf i}S^{+}_{\bf j}}\neq 0$ with broken TS 
\end{tabular}
\end{ruledtabular}
\end{center}
\end{table}
In TABLE \ref{table:correspondence}, 
the long-distance limit $|{\bf i}-{\bf j}|\rightarrow \infty$ 
is implied.  In the supersolid (SS) phase and its spin counterpart, 
translation symmetry is spontaneously broken (i.e. 
$\VEV{S^z_{\bf i}}$ and $\VEV{n_{\bf i}}$ modulate in space with 
non-trivial periods) as well as the axial $U(1)$.

In this paper, we reserve the terminology `NAF' 
for the Ising-like NAF phase 
and 
the corresponding phase in the Bose-Hubbard model (\ref{eqn:Bose-Hubbard}) 
will be called (checkerboard) `solid' or `half-filled solid'. 

\subsection{Spin wave Hamiltonian}
In order to rewrite the spin operators in terms of bosons, 
it is convenient to define magnons over a presumed reference state. 
In the case of NAF, all spins on the A-sublattice point upward and 
those on the B-sublattice downward (see Fig.\ref{Fig;config}).  
Therefore, it would be reasonable to introduce the following 
antiferromagnetic Dyson-Maleev (ADM) transformation
\cite{Dyson,Oguchi,Takahashi},
\begin{subequations}
\begin{align}
S_\boll^+&=\sqrt{2S}a_\boll,\ \ S_\boll^-=\sqrt{2S}a_\boll^\dagger\left(1-\frac{a_\boll^\dagger a_\boll}{2S}\right),\nonumber\\
&S^z_\boll=S-a^\dagger_\boll a_\boll,\ \ \ \text{for}\ \ \boll\in \text{A}\ .\\
S_\bolm^+&=\sqrt{2S}b_\bolm^\dagger,\ \ S_\bolm^-=\sqrt{2S}\left(1-\frac{b_\bolm^\dagger b_\bolm}{2S}\right)b_\bolm,\nonumber\\
&S^z_\bolm=-S+b^\dagger_\bolm b_\bolm,\ \ \ \text{for}\ \ \bolm\in \text{B}\ .
\end{align}
\label{eq;ADM}
\end{subequations}
If we introduce the Fourier transformation as
\begin{equation}
a_\boll=\frac{1}{\sqrt{N}}\sum_{\bf k} a_{\bf k}e^{i{\bf k}\cdot {\bf l}}\ ,\ \ b_\bolm=\frac{1}{\sqrt{N}}\sum_{\bf k} b_{\bf k}e^{i{\bf k}\cdot {\bf m}}\ ,
\end{equation}
($N$ is the number of sites of each sublattice), then the Hamiltonian
is given by
\begin{subequations}
\begin{align}
H=&H_0+H_1+\text{const}\ ,\label{Hbare}\\
H_0 =&\sum_{\bf k} S\left\{(\epsilon_0({\bf k})-h)a^\dagger_{\bf k} a_{\bf k}+(\epsilon_{0}({\bf k})+ h)b^\dagger_{\bf k} b_{\bf k} \right.\nonumber\\
&+\left.t_0({\bf k})(a_{\bf k} b_{-\bf k}+a_{\bf k}^\dagger b_{-\bf k}^\dagger)\right\} \ ,\label{H0bare}\\
\begin{split}
H_1 =&\frac{1}{N}\sum_{\bolk_1,\bolk_2,\bolq} \left\{ -2J_1^z C_1({\bf q}) a^\dagger_{\bolk_1+\bolq}b^\dagger_{\bolk_2-\bolq}a_{\bolk_1}b_{\bolk_2}\right.\\
+&(J_2^z C_2({\bf q})-J_2^\perp C_2({\bf \bolk_2}))a^\dagger_{\bolk_1+\bolq}a^\dagger_{\bolk_2-\bolq}a_{\bolk_1}a_{\bolk_2}  \\
+&(J_2^z C_2({\bf q})-J_2^\perp C_2({\bf \bolk_2-\bolq}))b^\dagger_{\bolk_1+\bolq}b^\dagger_{\bolk_2-\bolq}b_{\bolk_1}b_{\bolk_2}\\
-J_1^\perp&C_1({\bf \bolk_2})\left. (a^\dagger_{\bolk_1+\bolq}a_{\bolk_2-\bolq}^\dagger a_{\bolk_1}b_{-\bolk_2}^\dagger +b^\dagger_{\bolk_1+\bolq}b_{\bolk_2+\bolq}b_{\bolk_1}a_{-\bolk_2})\right\},\label{H1bare}
\end{split}
\end{align}
\label{boseHorigin}
\end{subequations}
where
\begin{subequations}
\begin{align}
\epsilon_0({\bf k})&=8J_1^z-6J_2^z+2J_2^\perp C_2({\bf k})\ ,\\
t_0({\bf k})&=2J_1^\perp C_1({\bf k})\ ,\\
C_1({\bf k})&=4\cos\frac{k_x}{2}\cos\frac{k_y}{2}\cos\frac{k_z}{2}\ ,\\
C_2({\bf k})&=\cos k_x+\cos k_y+\cos k_z\ .
\end{align}
\end{subequations}
Although this Hamiltonian is not hermitian and contains unphysical states\cite{Dyson}, we believe that the Hamiltonian given by eqs.(\ref{Hbare})-(\ref{H1bare}) correctly captures the low energy physics at and around the half-filled solid. 
Actually, in the case of magnon BEC just below the saturation field, though generally not proven, it is known for some specific models that the ferromagnetic Dyson-Maleev transformation, the Holstein-Primakoff transformation and the hard-core boson expansion for the $S=1/2$ case give the same ground state in a dilute bose gas approach\cite{Batyev,Nikuni-Shiba-2,Chubukov,HTUandKT}. 

\section{General Formalism}
\label{Sec;general}
In this section, we outline the dilute-Bose gas approach by which we shall investigate the SS phase around the half-filled solid appearing in the system described by the Hamiltonian eq.(\ref{boseHorigin}).

\subsection{Bogoliubov transformation}
In the following analysis, we frequently deal with Hamiltonians
of the following form:
\begin{equation}
\begin{split}
H_{\text{quad}}=S &  
\Bigl\{ 
(\epsilon({\bf k})-h)a^{\dagger}_{\bf k}a_{\bf k}
+(\epsilon({\bf k})+h)b^{\dagger}_{\bf k}b_{\bf k} 
\\
& +t({\bf k})(a_{\bf k}b_{-{\bf k}}+a_{\bf k}^{\dagger} 
b_{-{\bf k}}^{\dagger})\Bigr\} \; .
\end{split}
\label{HquadR}
\end{equation}
This is the most general quadratic Hamiltonian allowed by
hermiticity and sublattice symmetry. When we consider the
quadratic part (\ref{H0bare}) of the Hamiltonian $H$, the functions $\epsilon(\bolk)$
and $t(\bolk)$ should be taken as:
\begin{equation}
\epsilon({\bf k})=\epsilon_0({\bf k}),\ t({\bf k})=t_0({\bf k})\ ,\label{HS1quad}
\end{equation}
However, since the interaction $H_1$ shifts the grounds state, 
the renormalized quadratic Hamiltonian which leads to the exact Green's 
function including the self-energy do not in general 
coincide with (\ref{H0bare}).  
Generically the functions $\epsilon(\bolk)$
and $t(\bolk)$ are given by:
\begin{equation}
\epsilon({\bf k})=\epsilon_0({\bf k})+\epsilon^\prime({\bf k}),\ t({\bf k})=t_0({\bf k})+t^{\prime}({\bf k})\ ,\label{Generalet}
\end{equation}
where $\epsilon^\prime({\bf k})$ and $t^\prime({\bf k})$ are of the order of $S^{-1}$ since the interaction $H_1$ is of the order of $S^{0}$. 
In this paper, we approximately calculate the functions 
$\epsilon^\prime({\bf k})$ and $t^\prime({\bf k})$ 
in powers of $S^{-1}$ (Sec.~\ref{LargeS}) 
or of the Ising coupling constant $1/J_1^{z}$ 
(Sec.~\ref{LargeJ}).
Now let us assume that we have found an appropriate $H_{\rm quad}$.  
Then, in order to eliminate the off-diagonal terms 
$ab+a^\dagger b^\dagger$, 
we may introduce the following Bogoliubov transformation:
\begin{subequations}
\begin{align}
a_{\bolk} &=\cosh \theta_{\bf k} \alpha_{\bf k}-\sinh \theta_{\bf k} \beta_{\bf k}^\dagger\ ,\\
b_{\bf k} &=-\sinh \theta_{\bf k} \alpha_{\bf k}^\dagger+\cosh \theta_{\bf k} \beta_{\bf k}\ ,
\end{align}
\end{subequations}
which transforms $H_{\rm quad}$ to:
\begin{equation}
\begin{split}
H_{\rm quad}=S&\Bigl\{ (\epsilon_{\alpha}({\bf k})-h)\alpha^\dagger_{\bf k}\alpha_{\bf k}+(\epsilon_{\beta}({\bf k})+h)\beta^\dagger_{\bf k}\beta_{\bf k}\\
&+f({\bf k},\theta_{\bolk})(\alpha_{\bf k}\beta_{-\bf k}+\alpha_{\bf
 k}^\dagger\beta_{-\bf k}^\dagger)\Bigr\}\  . \\
\end{split}
\end{equation}
In the above, we have introduced two functions
\begin{subequations}
\begin{align}
&\epsilon_{\alpha}({\bf k})=\epsilon_{\beta}({\bf k}) \equiv\epsilon({\bf k})\cosh 2\theta_{\bf k}-t({\bf k})\sinh 2\theta_{\bf k}\ ,\label{epsilonAlpha}\\
&f({\bf k},\theta_{\bolk})\equiv-\epsilon({\bf k})\sinh 2\theta_{\bf k}+t({\bf k})\cosh 2\theta_{\bf k}
\end{align}
\label{sec2;Bog1}
\end{subequations}
If we choose $\theta_\bolk$ in such a way that $f(\bolk, \theta_\bolk) = 0$, 
i.e.
\begin{equation}
\tanh 2\theta_{\bolk}=\frac{t({\bf k})}{\epsilon({\bf k})}\label{sec2;Bog2}
\end{equation}
$H_{\rm quad}$ is diagonalized and reads
\begin{equation}
H_{\rm quad}=S(\epsilon_{\alpha}({\bf k})-h)\alpha^\dagger_{\bf k}\alpha_{\bf k}
+S(\epsilon_{\alpha}({\bf k})+h)\beta^\dagger_{\bf k}\beta_{\bf k}\ ,\label{diagH0}
\end{equation}
It is important to note that the magnetic field $h$ has different
signs for $\alpha$ and $\beta$. Assuming the (unique) minimum of the
spinwave excitation $\epsilon_{\alpha} (\bolk)$ takes place at $\bolk = \bolQ$, we may introduce the renormalized chemical potential by
\begin{equation}
\mu_{\rm \alpha} \equiv h-\epsilon_{\alpha}({\bf Q})\ .\label{muR}
\end{equation}
Now suppose we increase the external magnetic field $h$ (or $\mu_{\alpha}$). Then, the gap of the $\alpha$ $(\beta)$ boson decreases (increase) and the $\alpha$ bosons into an $\alpha$-SF phase discussed below.

\subsection{Supersolid from magnon-BEC}
In the previous subsection, we have seen that, as the external
magnetic field is increased, the $\alpha$ magnon condenses
at $\mu_{\alpha} = 0$ while the other remains gapped. Now, we show
that this BEC of the Bogoliubov-transformed magnons generally
leads to an SS phase. When a BEC occurs for
$\mu_{\alpha}\geq 0$, $\alpha_{\bolQ}$ takes a finite expectation value 
$\VEV{\alpha_{\bolQ}}\neq 0$ and,
correspondingly, the original bosons $a$, $b$ have the following
expectation values:\footnote{
In Sec.~\ref{Sec;general}, the hermiticity of the Hamiltonian is assumed.
In general, the non-hermitian DM Hamiltonian $H_{\rm DM}$ is given by $U^{-1}H_{\rm HP} U$, where $H_{\rm HP}$ is the Holstein-Primakoff transformed Hamiltonian and $U$ is the non-unitary operator, which recovers the hermiticity of $H_{\rm DM}$.\cite{Oguchi}
In NAF phase, we may not consider the effect of $U$ when we calculates the observables since the ground state (vacuum) is the eigenstate of $U$. If the boson condenses, $U$ may shift the vacuum and we must manipulate the operator $U$ explicitly. However, since we consider only a dilute-gas limit and need the observables obtained in the NAF phase (at $\mu_{\alpha}=0^{-}$), the hermiticity does not matter in the concrete discussion. 
}
\begin{equation}
\VEV{a_{{\bf Q}}}=\cosh \theta_{\bf Q}\VEV{\alpha_{{\bf Q}}},\ \ \ 
\VEV{b_{{\bf Q}}}=-\sinh \theta_{\bf Q}\VEV{\alpha_{{\bf Q}}^\dagger}\ .
\label{SSBEC}
\end{equation}
In a dilute-gas limit, when translated into the spin language, this implies the following spin configuration:
\footnote{
When there exist degenerate minima at several $\bolQ$s, there is a
possibility that the bosons at different $\bolQ$s simultaneously condense
and the magnetic structure may be different from the one
characterized by (\ref{SSconfig}). See appendix A for more detail.
}

\begin{subequations}
\begin{align}
&\VEV{S_\boll^x}=\sqrt{2S\rho}\cosh \theta_{{\bf Q}}\cos ({\bf Q}\cdot \boll+\varphi)(1+\frac{f(\Delta S)}{S}),\nonumber\\
&\VEV{S_\boll^y}=\pm\sqrt{2S\rho}\cosh \theta_{{\bf Q}}\sin ({\bf Q}\cdot\boll+\varphi)(1+\frac{f(\Delta S)}{S}),\nonumber\\
&\VEV{S^z_\boll}=(S-\Delta S) -\rho\cosh^2\theta_{{\bf Q}},\ \ \ \text{for}\ \ \boll\in \text{A}\ ,\\
&\VEV{S_\bolm^x}=-\sqrt{2S\rho}\sinh \theta_{{\bf Q}}\cos ({\bf Q}\cdot\bolm+\varphi)(1+\frac{f(\Delta S)}{S}),\nonumber\\
&\VEV{S_\bolm^y}=\mp\sqrt{2S\rho}\sinh \theta_{{\bf Q}}\sin ({\bf Q}\cdot\bolm+\varphi)(1+\frac{f(\Delta S)}{S}),\nonumber\\
&\VEV{S^z_\bolm}=-(S-\Delta S)+\rho\sinh^2\theta_{{\bf Q}},\ \ \ \text{for}\ \ \bolm\in \text{B}\ ,
\end{align}
\label{SSconfig}
\end{subequations}
where the real-space wavefunction is given by 
$\VEV{\alpha_{\bf r}}=\sqrt{\rho}\exp\{\pm i({\bf Q}\cdot {\bf r}+\varphi)\}$ 
and $\Delta S=1/N\sum_{q}\sinh^2\theta_q$. 
The function $f(\Delta S)=\Delta S/2+O(1/S)$ is obtained from 
the Holstein-Primakoff transformed operator $S^{\pm}$ and is independent of $\rho_{\bf Q}$ 
in the dilute-gas limit. One can easily see that this state may be thought of as an SS of magnons; 
an off-diagonal long-range (incommensurate) $xy$-order 
(which translates into an SF long-range order) and 
a diagonal (commensurate) 2-sublattice $z$-order coexist with each
other. 
In general, a modulation in the transverse component $S^{x,y}$ 
with the wave vector $\bolQ$ is incommensurate 
with the pattern of the $z$-order. 

If we denote the effective two-body interaction among the 
condensed bosons evaluated 
at $\mu_\alpha= 0^{-}$ 
by $\Gamma$, the leading term of the system energy 
is in general written, as a function of the condensate density $\rho$, as
\begin{equation}
\frac{E_{\text{eff}}}{N}\approx \text{const}+\frac{1}{2}\Gamma \rho^2-S\mu_{\alpha}\rho\ .\label{sysE}
\end{equation}
Then, provided $\Gamma > 0$, $\rho$ is given by minimizing $E$:
\begin{equation}
\frac{\rho}{S}=\frac{\mu_{\alpha}}{\Gamma},\ \ \ \text{for}\ \ \mu_{\alpha}\geq0\ .\label{onerho}
\end{equation}
However, the condition $\Gamma>0$ is {\it not} sufficient 
condition for the stability of the SS phase 
since there may be higher order terms with negative coefficients 
in $E_{\text{eff}}$, which  may select a very large 
value of $\rho$ and eventually destabilize the SS phase. 
If $\Gamma\leq 0$, on the other hand, one may expect 
a phase separation accompanied by magnetization jump 
near $\mu_{\alpha} = 0$. 
For both cases, there exists an additional possibility 
of more exotic phases where single-particle BECs are 
no longer relevant.\cite{HTUandKT} 

The low-energy excitation spectrum of the SS phase is easily 
obtained as in the ordinary superfluid Bose gas\cite{HuaShi}. 
Defining $\bolk\equiv \bolq{-}\bolQ$, we may expand 
$\epsilon_{\alpha}(\bolq)=\epsilon_{\text{min}}+k_i k_j/(2m_{ij})+\cdots $, 
where the summation over repeated indices is implied.  
We can diagonalize $m_{ij}$ to obtain a standard dispersion 
$k_i k_j/2m_{ij}=k_i^{\prime 2}/(2m^\prime_{i})\equiv 
\epsilon_{\text{g}}(\bolk^\prime)$. 
Using this notation, the excitation spectrum of the SS phase is given by 
\begin{equation}
\Omega_{\text{SS}}(\bolk)=
\sqrt{\epsilon_{\text{g}}(\bolk)^{2}+2S\mu_{\rm R}\epsilon_{\text{g}}(\bolk)}\approx 
\sqrt{2S\mu_{\rm R}\epsilon_{\text{g}}(\bolk)}.
\end{equation}

For finite temperature, the Bose condensed bosons are suppressed, 
and the critical temperature is given by
\begin{equation}
k_{\rm B} T_{\rm c} =2.087(m_xm_ym_z)^{-\frac{1}{3}}(\frac{S\mu_{\alpha}}{\Gamma})^{\frac{2}{3}}\ .
\label{kT}
\end{equation}
For $T>T_c$, the long-range order disappears and $\VEV{S^{\pm}}=0$.

Above discussions assume the dilute-gas limit, 
where the scattering length is much smaller than the average 
inter-atomic distance $\rho^{-1/3}$. Specifically, 
our approximation is valid when 
\begin{equation}
\Gamma (m_xm_ym_z\rho)^{1/3} \ll 1\ .
\end{equation}

To summarize, the knowledge about the wave number $\bolQ$ 
at which the magnon BEC occurs, the effective mass $m_i$
and the effective (2-body) interaction $\Gamma$ for the condensed 
bosons enables us to derive the stability, 
the spin configuration which is not commensurate 
with the assumed sublattice structure, the quasi-particle 
excitation spectrum and the critical temperature of the SS 
phase. Therefore, the analysis boils down to the calculation 
of $\bolQ$ and $\Gamma$. A remark is in order here about the definition 
of the bosonic vacuum. In eq.(\ref{eq;ADM}), it is implicitly assumed 
that the NAF phase gives a well-defined vacuum (i.e.
the ground state when the condensate is absent) for the two 
bosons. 
In general, the NAF state shown in Fig.\ref{Fig;config} 
suffers from quantum fluctuations and the above assumption 
is justified either for the semiclassical (i.e. large-$S$) case 
or the Ising-like (i.e. large-$J^z_1 /J_1^\perp$) limit\footnote{%
The analysis near the saturation field is free from this problem; 
the fully polarized state is an exact eigenstate and high enough 
magnetic field guarantees the validity of the reference state. 
}

In the following sections, we carry out the calculation by combining 
the ladder approximation with the large-$S$ and the Ising expansions. 
Concretely, in Sec.~\ref{LargeS}, we will obtain eq.(\ref{Generalet}) 
and the interaction $\Gamma$ by the large-$S$ expansion 
up to the second order in $S^{-1}$. 
At the first order, our approach will reproduce the results of the MFT; 
there are three types of SSs. 
At the second-order perturbation, quantum fluctuations may change 
the properties of the solid and the SSs qualitatively. 
However, we will see that the large-$S$ expansion is not reliable 
to calculate $\Gamma$ when the Ising-like anisotropy $J_1^z$ is large. 
To overcome this difficulty, we will study eq.(\ref{Generalet}) 
and $\Gamma$ by the Ising expansion up to the second order in Sec.~\ref{LargeJ}. 
At the first order, quantum fluctuations 
 suppress the interactions, 
but the stability of the SS itself is known by the MFT. 
Although we will find the stable bound-magnon state, 
this condensed phase may be phase-separated for large $J_1^z$. 
In the second order, we will see the effect of quantum fluctuations 
on the stability of SS clearly. 

\section{perturbation theory in $S^{-1}$}\label{LargeS}
In this section, we study the physics of the SS phase by the 
perturbation theory in the parameter $S^{-1}$.
The first-order calculation gives the same ground-state 
phases as the MFT. 
At the second order, on the other hand, 
quantum fluctuations play 
an important role and may destroy the classically stable solid (NAF) 
or the SS phase. 

\subsection{First-order perturbation}\label{sec;S1}
If we assume $\theta_{\bf k}$ by
\begin{equation}
\tanh 2\theta_{\bf k}^{(1)}=\frac{t_0({\bf k})}{\epsilon_0({\bf k})}=\frac{J_1^\perp C_1({\bf k})}{4J_1^z-3J_2^z+J_2^\perp C_2({\bf k})}\ ,\label{BogoC}
\end{equation}
the quadratic part of the Hamiltonian is diagonalized up to $O(S)$ (see eq.(\ref{Generalet})). We note that $\theta_k^{(1)}$ is well-defined when $|\tanh 2\theta_{\bf k}^{(1)}| \leq 1$. Concretely, the half-filled solid is stable at $h=0$ when
\begin{equation}
4J_1^z-3J_2^z+3J_2^\perp \geq 4|J_1^\perp|\ .\label{PlateauCondition}
\end{equation}
If this inequality is not satisfied, the spins align in the $xy$-plane (SF) (see eqs.(\ref{EIsing}) and (\ref{ESF})). 
Meanwhile, even when the classical ground state is CAF ($2J_1^z<3J_2^z$), this inequality may be satisfied and then the metastable NAF phase against the one magnon fluctuation may be obtained. In this paper, we will not discuss the CAF case any more.

\begin{figure}[H]
\begin{center}
\includegraphics[scale=1]{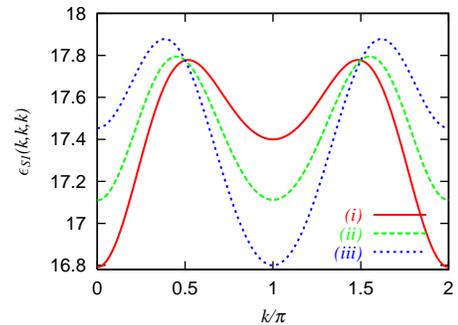}
\caption{(color online) The dispersion relation of the excitation energy $\epsilon_{\rmS 1}(\bolk)$ for $\bolk=(k,k,k),\ h=0,\ J^z_1=3,\ J^z_2=1,\ J_1^\perp=-1$. (i) is obtained at $J_2^\perp =0.1$ and the minimum is at $\bolk=\bolQ_1$. 
(ii) is at $J_2^\perp =0.148$ and the minima are at both $\bolk=\bolQ_1$ and $\bolQ_2$. 
(iii) is at $J_2^\perp =0.2$ and the minimum is at $\bolk=\bolQ_2$. 
\label{Fig;S1disp}
}
\end{center}
\end{figure}

Let us discuss the minimum of the dispersion $\epsilon_{\alpha}({\bf k})=\epsilon_{\rmS 1}^{(1)}({\bf k})$ (see eq.(\ref{epsilonAlpha})) to determine the structure of the SS. 
From (\ref{BogoC}), the dispersion relation reads,
\begin{equation}
\begin{split}
\epsilon_{\rmS 1}^{(1)}({\bf k})&=\sqrt{\epsilon_0(\bolk)^2-t_0(\bolk)^2}\\
&=\sqrt{(8J^z_1-6J^z_2+2J_2^\perp C_2({\bf k}))^2-4J_1^{\perp2} C_1({\bf k})^2}\ .
\end{split}
\end{equation}
where the superscript $(i)$ of $\epsilon_{\rmS 1}^{(i)}(\bolk)$ denotes that the function is evaluated at $\theta_k=\theta_k^{(i)}$. In the following we shall use this notation to the other arbitral functions of $\theta_{\bolk}$. 
The minimum is obtained by setting ${\bf {\bf Q}}={\bf {\bf Q}}_1=(0,0,0)$ or ${\bf {\bf Q}}_2=(\pi,\pi,\pi)$. Although we can not exclude other possibilities generally, this is always the case for the parameter sets used in this paper. 
It is convenient to introduce $\Lambda$ as
\begin{equation}
\Lambda\equiv\epsilon_{\rmS 1}^{(1)}({\bf {\bf Q}}_2)^2-\epsilon_{\rmS 1}^{(1)}({\bf {\bf Q}}_1)^2=16\bigl(-J_2^\perp(12J_1^z-9J_2^z)+4J_1^{\perp2}\bigr)\ ,\label{Lambda}
\end{equation}
Then, one chooses $\bolk={\bf Q}_1$ when $\Lambda>0$ or $\bolk={\bf Q}_2$ when $\Lambda<0$. 
We have checked that the SS with ${\bf Q}_1$ (SS1) is always favored for $J_2^\perp\leq0$. 
And when the Ising-like anisotropy is large, i.e., $12J^z_1-9J^z_2\gg 4J_1^{\perp}$, very small positive $J_2^\perp$ selects the SS with ${\bf Q}_2$ (SS2). 
For each case, we plot the dispersion relation $\epsilon_{\rmS 1}(\bolk)$ along the 
$(1,1,1)$-direction in Fig.\ref{Fig;S1disp}. 

From eq.(\ref{muR}), the chemical potentials of both phases are given by 
\begin{equation}
\mu_{{\rmS 1}{\rm SS}i} \equiv h-\epsilon_{\rmS 1}({\bf Q}_i)\ , \ \ {\rm for}\ i=\{1,2\}\ ,\label{muS1}
\end{equation}
where the subscript `S$n$' means that the interactions are expanded 
up to $n$-th order in $S^{-1}$  and `SS$i$' represents the types 
of the SS.
The effective masses are isotropic and are respectively given by
\begin{subequations}
\begin{align}
m_{{\rmS 1}{\rm SS1}}&=\frac{\sqrt{(4 J_1^z-3 J_2^z+3 J_2^\perp)^2-16
   J_1^{\perp 2}}}{2S\left\{-J_2^\perp (4 J_1^z-3
   J_2^z+3 J_2^\perp)+4 J_1^{\perp 2}\right\}}\ ,\\
m_{{\rmS 1}{\rm SS2}}&=\frac{1}{2SJ_2^\perp}\ .
\end{align}
\label{masscl}
\end{subequations}

When $\Lambda=0$, the two minima are degenerate and we have to take into account two independent condensates and phases which are not characterized by (\ref{diagH0}) may appear. A brief discussion on this case is given in appendix \ref{AP;SS3}. The SS phase of 4-sublattice structure (SS3) actually exists for certain parameter sets. 
There exist three types of SS around the half-filled solid. 

Next, we consider the stability of the SS phase. The 2-body interaction between $\alpha$ bosons is given by the first-order diagram since the bare Green's function of $\alpha$ ($\beta$) bosons is $i/(\omega-S\epsilon^{(1)}_\alpha(\bolk)\pm h)=O(S^{-1})$ for $\omega\sim -\mu_\alpha$ and the vertex function is $O(S^0)$. The alternative view is that, if we rescale the Hamiltonian by $S^{-1}$, the vertex function is $O(S^{-1})$ and the diagram is suppressed by $S^{-1}$ for each vertex. 
Therefore, we need only the vertex function between $\alpha$ bosons. By replacing $a_{\bf k}\rightarrow \cosh \theta^{(1)}_{\bf k}\alpha_{\bf k}$ and $b_{\bf k}\rightarrow -\sinh \theta^{(1)}_{\bf k} \alpha_{\bf k}^\dagger$ in $H$, the interaction term of $\alpha$ bosons appears as the following form:
\begin{equation}
\frac{1}{2N}\sum V_{\alpha}({\bf q};{\bf k}_1,{\bf k}_2)\alpha_{\bolk_1+\bolq}^\dagger\alpha_{\bolk_2-\bolq}^\dagger\alpha_{\bolk_1}\alpha_{\bolk_2}\ ,
\end{equation}
where the factor 2 in front of N is considered for the symmetry factor. 
For the case $\Lambda>0$ and ${\bf Q}={\bf Q}_1(=0)$, $\Gamma$ is given by,
\begin{equation}
\Gamma_{{\rm S1SS1}}=V_{\alpha}(0;\bolQ_1,\bolQ_1)=6(J^z_2-J_2^\perp)\ ,\label{S1GammaSS1}
\end{equation}
Thus, SS phase of $\bolQ_1$ is stable for $J^z_2-J_2^\perp>0$.

For the case $\Lambda<0$ and ${\bf Q}={\bf Q}_2$, $\Gamma$ is given by,
\begin{equation}
\Gamma_{{\rm S1SS2}}=V_{\alpha}(0;\bolQ_2,\bolQ_2)=6(J^z_2+J_2^\perp)\ ,\label{S1GammaSS2}
\end{equation}
In this phase, the Spin on $B$-sublattice does not have the transverse magnetization even for $\mu_{\rm S1SS2}>0$ since $\sinh \theta^{(1)}_{\bolQ_2}=0$. 

\begin{figure}[H]
\begin{center}
\includegraphics[scale=1.2]{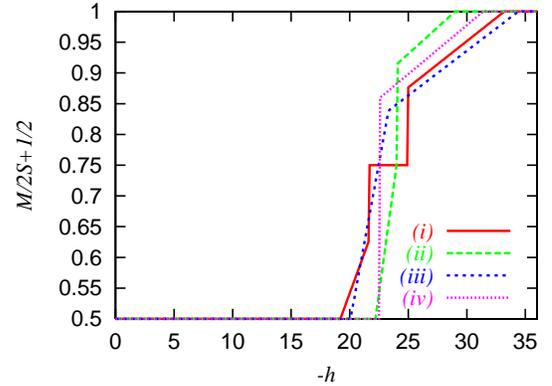}
\caption{(color online) Magnetization curves obtained from the MFT for $J^z_1=3,\ J_1^\perp=-1$. We assume 2-sublattice structure or the quarter-filled solid. $M$ is given by $(1/2N)\sum_\boll^{2N} \VEV{S^z_\boll}$ . (i) curve is obtained for $J^z_2=0.5,\ J_2^\perp=0.3$. (ii) is for $J^z_2=-0.1,\ J_2^\perp=0.4$. (iii) is for $\ J^z_2=0,\ J_2^\perp=-0.4$. (iv) is for $J^z_2=-0.5,\ J_2^\perp=-0.4$. All curves has a half-filled solid around $h=0$ and SF phase below the saturation field. (i) curve has SS2 just ahead the half-filled solid, and the quarter-filled solid. (ii) has SS2 also. (iii) has SS1 which is connected to SF phase continuously. (iv) does not have SS phase.
\label{Fig;jiba1}}
\end{center}
\end{figure}

To see the validity of the above picture, we compare the above result with that of the MFT. Although there exists the extensive MFT calculation of this model for\cite{Matsuda-Tsuneto,Liu-Fisher} $J_2^\perp <0$, to the best of our knowledge, there is not the appropriate mean field calculation of the models for $J_2^\perp>0$. Hence, we redo the MFT for $S=1/2$. Now, the ground state energy is obtained by replacing the operators in $H$ with their expectation values of Pauli matrices on each site, e.g., $\sum_{\VEV{i,j}}S_i^zS_j^z\rightarrow \sum_{\VEV{i,j}}S^2 \VEV{\sigma_i^z}\VEV{\sigma_j^z}$. We compare energies of the three types of spin configurations, 
\begin{subequations}
\begin{align}
\begin{split}
&\frac{E_{1}^\text{mean}}{NS^2}=8J^z_1\VEV{\sigma^z}\VEV{\sigma^z}^\prime+3J^z_2(\VEV{\sigma^z}^2+\VEV{\sigma^z}^{\prime2})\\
&\hspace{0.5cm}-8|J_1^\perp|\tau\tau^\prime+3J_2^\perp(\tau^2+\tau^{\prime2})+h(\VEV{\sigma^z}+\VEV{\sigma^z}^\prime) ,
\end{split}\\
\begin{split}
&\frac{E_{2}^\text{mean}}{NS^2}=8J^z_1\VEV{\sigma^z}\VEV{\sigma^z}^\prime+3J^z_2(\VEV{\sigma^z}^2+\VEV{\sigma^z}^{\prime2})\\
&\hspace{0.7cm}-3J_2^\perp(\tau^2+\tau^{\prime2})+h(\VEV{\sigma^z}+\VEV{\sigma^z}^\prime)\ ,
\end{split}\\
&\frac{E_{1/4\text{filled}}}{NS^2}=-h\ .
\end{align}
\end{subequations}
where $\tau=\sqrt{\VEV{\sigma^x}^2+\VEV{\sigma^y}^2}=\sqrt{1-\VEV{\sigma^z}^2}$. $E_{1}^\text{mean}$ is obtained from the 2-sublattice structure, $E_{2}^\text{mean}$ is from the 2-sublattice structure of $\VEV{\sigma^z}$ and $\tau$ with AF-$(\pi,\pi,\pi)$ $\VEV{\sigma^{x,y}}$ ordering on each sublattice, and $E_{1/4\text{filled}}$ is from the quarter-filled solid. In this paper, we ignore the possibility that another types of SS phases appear around the quarter-filled solid as in the model on the square lattice\cite{Pich-Frey,Dang}. By minimizing each energy numerically, we obtain magnetization curves for various parameters. We confirmed that the Bose-gas approach gives the same results as the MFT one. The specific examples are shown in FIG.\ref{Fig;jiba1}.

\subsection{Second-order perturbation}
\label{sec:secondorder}
To see the effect of quantum fluctuation more clearly, 
we consider the second-order perturbation theory in the parameter $S^{-1}$. 
In this order, the ground-state phase may become different from the mean-field one. 

To begin with, 
let us consider the state of the half-filled solid by diagonalizing the quadratic term in the Hamiltonian. 
When the interaction terms (\ref{H1bare}) are written in terms of the Bogoliubov-transformed bosons and put into the normal-order form, additional quadratic terms appear. 
As a result, the quadratic part of Hamiltonian reads,
\begin{equation}
\begin{split}
H_{\rm quadS2}=&\Bigl\{(S\epsilon_{0}({\bf k})-T_1(\bolk))\cosh 2\theta_{\bolk}\\
&-(St_0(\bolk)-T_2(\bolk))\sinh 2\theta_{\bolk}-Sh\Bigr\}\alpha^\dagger_{\bf k}\alpha_{\bf k}\\
+&\Bigl\{(S\epsilon_{0}({\bf k})-T_1(\bolk))\cosh 2\theta_{\bolk}\\
&-(St_0(\bolk)-T_2(\bolk))\sinh 2\theta_{\bolk}+Sh\Bigr\}\beta^\dagger_{\bf k}\beta_{\bf k}\\
+\Bigl\{&-(S\epsilon_{0}({\bf k})-T_1(\bolk))\sinh 2\theta_{\bf k}\\
+&(St_0(\bolk)-T_2(\bolk))\cosh 2\theta_{\bf k}) \Bigr\}(\alpha_{\bf k}\beta_{\bf k}+\alpha_{\bf k}^\dagger\beta_{\bf k}^\dagger)\ ,
\end{split}
\label{Hqfull}
\end{equation}
where $T_k$s are given by eq.(\ref{SpinT}). 
Even in the normal-ordered 2-body interaction terms, there exists the terms which shift the vacuum with respect to $\alpha$ and $\beta$ (e.g. $\alpha^\dagger\beta^\dagger\alpha^\dagger\beta^\dagger\ket{0}\neq 0$), which leads to the self energy. However, this contributes the Green's function in the third-order of $S^{-1}$ and we neglect the self energy in our approximation. 
We note that, even if we use Holstein-Primakoff transformation, the same quadratic Hamiltonian is obtained up to the second order in $S^{-1}$. The difference between the two boson representations (i.e. Dyson-Maleev and Holstein-Primakoff) 
appears in the 2-body interaction term. 

Now, $\theta_\bolk$ is given by solving
\begin{equation}
\begin{split}
&(-S\epsilon_0({\bf k})+T_1(\bolk))\sinh 2\theta_{\bf k}\\
&+(St_0({\bf k})-T_2(\bolk))\cosh 2\theta_{\bf k}=0\ .
\end{split}
\end{equation}
To evaluate $T_{1,2}$, we need the explicit form of the function $\theta_{\bolk}$. Since $T_{1,2}$ is suppressed by a factor $1/S$ in the diagonalization procedure, we use $\theta_{\bolk}^{(1)}$ 
which is obtained in the first-order calculation 
{for the integrands in eq.(\ref{SpinT}). 
Therefore, $\theta_k^{(2)}$ which is corrected up to second order is given by
\begin{equation}
\tanh 2\theta_k^{(2)}=\frac{t_0({\bf k})-T^{(1)}_2(\bolk)/S}{\epsilon_0({\bf k})-T^{(1)}_1(\bolk)/S}\ .\label{S2tanh}
\end{equation}
If $|\tanh 2\theta_k^{(2)}|> 1$, the spinwave expansion concludes
that the half-filled solid is unstable and that other phases may take over. In fact, this happens for certain choices of the parameters. The detailed result will be discussed in sec.\ref{sec:lattice}.
Then the quadratic Hamiltonian and the dispersion relation $\epsilon_{\rm S2}(\bolk)$ are given respectively by 
\begin{equation}
\begin{split}
H_0^\prime=S\left(\epsilon_{\text{S2}}({\bf k})-h \right)\alpha^\dagger_{\bf k}\alpha_{\bf k}
+S\left(\epsilon_{\text{S2}}({\bf k})+h \right)\beta^\dagger_{\bf k}\beta_{\bf k}\ ,
\end{split}
\label{Hqfull}
\end{equation}
\begin{equation}
\epsilon_{\text{S2}}({\bf k})
=\sqrt{\left(\epsilon_0(\bolk)-\frac{T_1^{(1)}(\bolk)}{S}\right)^2
-\left( t_0(\bolk)-\frac{T_2^{(1)}(\bolk)}{S}\right)^{2}}\ .
\end{equation}
If we introduce the appropriate constants $a_1,\ldots ,a_3$, the above phonon dispersion $\epsilon_{\text{S2}}(\bolk)$ may be written generally as:
\begin{equation}
\epsilon_{\text{S2}}(\bolk)=\sqrt{(a_1+a_2C_2(\bolk))^2-(a_3C_1(\bolk))^2}\ ,
\end{equation}
and qualitatively the same dependence on $\bolk$ as in the first-order case is obtained. In our calculations, the minimum is always locked at $\bolQ_{1}=(0,0,0)$ or $\bolQ_{2}=(\pi,\pi,\pi)$, which respectively corresponds to SS1 or SS2. 
The criterion, which determines the structure and the effective mass for each phase, is easily obtained in the same manner as in the first-order case (see eq.(\ref{Lambda}) and (\ref{masscl})). However, the explicit forms are somewhat lengthy and we do not show them in this paper. In the following, we shall concentrate on the physics of SS1 and SS2 and shall not discuss SS3 further. The chemical potential $\mu_{{\alpha}}$, which controls the onset of BEC, are also different from the first-order one (\ref{muS1}) and is given by 
\begin{equation}
\mu_{{\rmS 2}{\rm SS}i} \equiv h-\epsilon_{\rmS 2}({\bf Q}_i)\ , \ \ {\rm for}\ i=\{1,2\}\ .
\end{equation}

Next, we briefly recapitulate the method by which we calculate the effective interaction $\Gamma$ among the condensed bosons. 
We simply evaluate the diagrams up to the second order in $S^{-1}$.
We have one diagram at the first order and six at the second order. The second-order diagrams are shown in Fig.~\ref{Fig;spin2}. 
To evaluate the second order diagram, we use the bare Green's function at $\mu_{{\rmS 2}{\rm SS}i}=0$ in the dilute bose gas approximation:
\begin{equation}
\VEV{T(\alpha_k \alpha_k^\dagger)(\omega)}=\frac{i}{\omega-S\left(\epsilon_{\text{S2}}({\bf k})-\epsilon_{\text{S2}}(\bolQ_i)\right)+i0^+}\ .
\end{equation}
In the presence of a finite condensate $|\VEV{\alpha} |^2=\rho\propto \mu$, the Green's function, which is obtained for a new operator $\alpha^\prime=\alpha-\VEV{\alpha}$, gets modified continuously from the one at the onset of
BEC\cite{HuaShi}. Specifically, $\VEV{\alpha^\prime\alpha^{\prime\dagger}}_{{\rm at}\ \mu > 0}=\VEV{\alpha\alpha^\dagger}_{{\rm at}\ \mu=0}+O(\mu)\ $, and $\VEV{\alpha^\prime\alpha^{\prime}}_{{\rm at}\ \mu > 0}=O(\mu)$.
In short, the modified quadratic Hamiltonian and the effective interaction $\Gamma$ calculated above tell us the stability and the low-energy physics of solid and SS phase. The detailed results are shown in Sec.~\ref{sec:lattice}.

\begin{figure}[H]
\begin{center}
\includegraphics[scale=0.4]{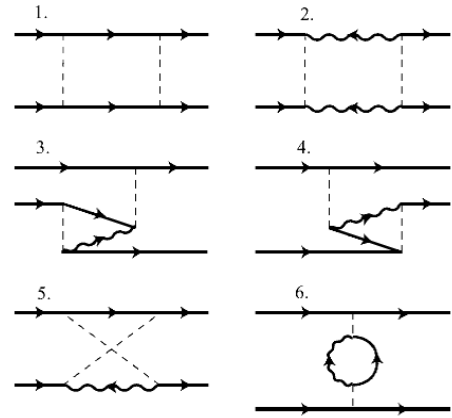}
\caption{The second-order (one-loop) diagrams in $S^{-1}$. Straight lines (wavy lines) denote $\alpha\ (\beta)$ bosons. Broken lines denote the momentum transfer at the interaction.
\label{Fig;spin2}}
\end{center}
\end{figure}

Finally, we consider the validity of the expansion 
of the exponential in powers of the interaction terms 
in the path integral when we calculate $\Gamma$. 
If we were able to take into account an infinite number of terms, the expansion would be correct. However, now we sum up only a finite number of terms. Thus, we need a criterion, even though naive, for determining the reliability of the expansion. A natural candidate may be the magnitude of the expanded interaction terms. To see this explicitly, we consider the following simple boson model on the simple cubic lattice:
\begin{equation}
H_{s}=\sum_k \frac{{\bolk}^2}{2m}d^\dagger_{\bolk} d_\bolk + \frac{1}{2N}\sum_{\bolk_1,\bolk_2,\bolq} 2\lambda d^\dagger_{\bolk_1+\bolq}d^\dagger_{\bolk_2-\bolq}d_{\bolk_1}d_{\bolk_2}
\end{equation}
In this model, the low-energy effective interaction $\Gamma_s$ between the condensed bosons is exactly obtained as $\Gamma_s=2\lambda /(1+(2/\pi)m\lambda)=2\lambda \sum_n (-(2/\pi)m\lambda)^n$. 
The dimensionless constant $m\lambda$ captures the magnitude of the expanded interaction terms. Thus, for general lattice boson models, we may expect that (mass)$\times$(coupling constant) gives a simple criterion for the validity of the expansion. 

Let us apply the above criterion to our case. 
For the boson masses, we use (\ref{masscl}), which are correct up to the first order in $S^{-1}$, for simplicity. 
An appropriate choice of the coupling constants may be $J_i^z$ and $J_i^\perp$ for $i=\{1,2\}$. 
For SS2, the criterion reads $J_i^z m_{2{\rm cl}}=J_i^z/(2SJ_2^\perp)$. Hence, however large the spin $S$ may be, the series expansion of $\Gamma$ eventually
diverges for relatively large Ising anisotropy. 
Similarly for SS1, the perturbation expansion is not converging for large
Ising anisotropy since $m_{1{\rm cl}}\sim -1/(2SJ_2^\perp)$. 
We have one more problem in the evaluation of $\Gamma$; when the energy dispersion at the solid is nearly gapless (i.e., $\tanh 2\theta_{\bolk=0}\approx 1$ in eqs.(\ref{sec2;Bog1}) and (\ref{sec2;Bog2})), $\cosh 2\theta_0$ and $\sinh 2\theta_0$ have large values(for $\tanh 2\theta\rightarrow 1$, $\theta \rightarrow \infty$). 
We note that these problems are peculiar to the evaluation of $\Gamma$ and the low-energy physics of the solid (NAF) is well understood by the large-$S$ expansion. 

From the above discussion, we may conclude
that the SS phases obtained within the MFT, which do not
change even after the first-order 1/S-correction is taken into
account, might be destroyed at higher orders by quantum fluctuations.
Since the perturbation expansion described above is
ill-behaved for large Ising anisotropy, we have to take another
approach to closely investigate the fate of the SS phases. In
the next section, we shall introduce another perturbation theory
with respect to large Ising anisotropy. A reliable treatment
of $\Gamma$ for the case with $\tanh 2\theta_{\bolk=0}\approx 1$ remains to be an open problem.


\section{perturbation theory in large Ising-like anisotropy}\label{LargeJ}
In the limit $J_1^z\nearrow \infty$, the system behaves like the Ising model. 
In this section, we compute $\Gamma$ by the perturbation theory in $(J_1^{z})^{-1}$. 
Specifically, we develop an expansion in small coupling constants $(J_2^z,J_1^\perp,J_2^\perp)$.

\subsection{First-order perturbation}
If we diagonalize the bare quadratic Hamiltonian $H_0$ (\ref{H0bare}), 
\begin{equation}
\tanh 2\theta_{\bf k}=\frac{J_1^\perp C_1({\bf k})}{4J_1^z-3J_2^z+J_2^\perp C_2({\bf k})}= O(1/J_1^{z})\ ,
\end{equation}
Then, 
\begin{equation}
\cosh \theta_\bolk=1+O((J_1^{z})^{-2})\ ,\ \ \sinh \theta_\bolk=O((J_1^{z})^{-1})\ .\label{Isingtheta1}
\end{equation}
If we assume that the exact $\theta_{\bolk}$ obtained by eq.(\ref{sec2;Bog2}) has the same property, the self-energy contribution to the quadratic Hamiltonian ($\epsilon^\prime(\bolk),\ t^\prime (\bolk)$ in (\ref{Generalet})) is up to $O((J_1^z)^0)$ and the dependence on $J_1^z$ of $\theta_\bolk$ is maintained as (\ref{Isingtheta1}).
Therefore, the leading-order Hamiltonian in $J_1^z$ reads
\begin{equation}
\begin{split}
H_{\rm I1}&=
\sum_\bolk S(\epsilon_0({\bf k})-h)\alpha^\dagger_\bolk\alpha_\bolk
+\sum_\bolk S(\epsilon_0({\bf k})+h)\beta^\dagger_\bolk\beta_\bolk\ ,\\
+&\frac{1}{N}\sum_{\bolk_1,\bolk_2,\bolq} (J_2^z C_2({\bf q})-J_2^\perp C_2({\bf \bolk_2}))\alpha^\dagger_{\bolk_1+\bolq}\alpha^\dagger_{\bolk_2-\bolq}\alpha_{\bolk_1}\alpha_{\bolk_2} \ .
\label{HIsing1}
\end{split}
\end{equation}
where we neglect the 2-body interaction term containing $\beta$ bosons
since the gap of $\beta$ boson is $O(J_1^{z})$ when the gap 
of $\alpha$ boson closes. 
The meaning of the subscript `I$n$' is similar to that of 
`S$n$' in the previous section; it means that terms are kept up to $n$-th order in the Ising expansion.
The minimum of the dispersion is obtained at ${\bf {\bf Q}}_1=(0,0,0)$ for $J_2^{\perp}<0$ (SS1) or ${\bf {\bf Q}}_2=(\pi,\pi,\pi)$ for $J_2^{\perp}>0$ (SS2). The chemical potential and the effective mass are respectively given by
\begin{subequations}
\begin{align}
\mu_{\rm I1}&=h-(8J_1^z-6J_2^z-6|J_2^\perp|)\ ,\label{muI1} \\
m_{\rm I1}&=\frac{1}{2S|J_2^\perp|}\ .
\end{align}
\end{subequations}

Next, let us evaluate the interaction $\Gamma$ among the $\alpha$ bosons.
Since both the Green's function of $\alpha$ bosons and the coupling constants of interaction are $O((J_1^{z})^0)$, the all-order diagrams equally contribute to $\Gamma$, which is given by the ladder diagram (Fig.\ref{Fig;ladder}). The ladder diagram $T$ evaluated at the solid satisfies
\begin{multline}
T(\bolq;\bolk_1,\bolk_2)  =K(\bolq;\bolk_1,\bolk_2)\\
- \frac{1}{N}\sum_{\bolq^\prime}\frac{T(\bolq^\prime;\bolk_1,\bolk_2)
K(\bolq-\bolq^\prime;\bolk_1+\bolq,\bolk_2-\bolq)}{\omega(\bolk_1+\bolq^\prime)
+\omega(\bolk_2-\bolq^\prime)-\omega(\bolk_1)-\omega(\bolk_2)}. 
\label{laddereq}
\end{multline}
where the particle corresponding to the external line is assumed to be 
a real one with the energy $\omega(\bolk)-\mu$. One obtains the parameter $\Gamma$ for SS$_i$ ($\bolQ_i$) as $\Gamma_{{\rm SS}_i}=T(0,\bolQ_i,\bolQ_i)$. 

\begin{figure}[H]
\begin{center}
\includegraphics[scale=0.4]{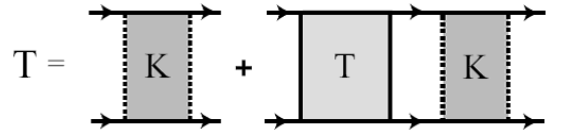}
\caption{Ladder diagram: $T$ represents the ladder diagram and $K$ represents the kernel which is not reducible to the product of 2-perticle Green's function.  
\label{Fig;ladder}}
\end{center}
\end{figure}

Now, the kernel $K$ and the energy $\omega$ are given by
\begin{subequations}
\begin{align}
K_{\rm I1}(\bolq;\bolk_1;\bolk_2)&=2J_2^z C_2(\bolq)-J_2^\perp(C_2(\bolk_1)+ C_2(\bolk_2)),\\
\omega_{\rm I1}(\bolk)&=2S(J_2^\perp C_2(\bolk)+3|J_2^\perp|)\ .
\end{align}
\end{subequations}
The self-consistent equation (\ref{laddereq}) for the ladder diagram is exactly solvable\cite{Batyev,Nikuni-Shiba-2} and we obtain
\begin{equation}
\Gamma_{{\rm
 I1SS}_i}=\frac{6(J_2^z+|J_2^\perp|)}{1+0.258\frac{(J_2^z+|J_2^\perp|)}{S|J_2^\perp|}}\
 ,
\label{eqn:GammaI1Qi}
\end{equation}
for $i=\{1,2\}$. If the limit $S\rightarrow \infty$ is taken, $\Gamma$ reduces to the first-order result in the large-S expansion ((\ref{S1GammaSS1}) and (\ref{S1GammaSS2})). For finite $S$, $\Gamma$ is suppressed by quantum fluctuations. 
However, concerning the stability of the SS phases, there is no difference from the MFT result as far as the denominator is
positive. We note that if 
\begin{equation}
1+0.258\frac{(J_2^z+|J_2^\perp|)}{S|J_2^\perp|}=0\ ,
\end{equation}
the effective interaction $\Gamma_{{\rm
 I1SS}_i}$ diverges. 
This suggests the possibility of the SS accompanied by the bound-magnon BEC. 
A brief discussion on this issue 
will be given in Sec.~\ref{2magBEC}.

\subsection{Second-order perturbation}
\label{sec:Ising-2}
In this section, we study the SS phase around the solid in the second-order perturbation in $(J_1^{z})^{-1}$ with the help of large $S$ expansion. 

To obtain the renormalized quadratic Hamiltonian, we perform the Bogoliubov transformation and normal-order the interaction term, 
sorting out the terms based on (\ref{Isingtheta1}).
Since near the boundary of NAF-CAF transition quantum fluctuation may play an important role, we keep only terms of order $O(J_2^z/(J_1^{z})^{2})$ (we neglect the order $O((J_2^z/J_1^{z})^{n}/J_1^z)$ terms with $n\geq2$). Even on the classical boundary, $6J_2^z=(1/2)8J_1^z$ and the $J_2^z$ times coordination number is suppressed by the large-anisotropy $J_1^z$. Therefore, the expansion may work.
Since the renormalized $\theta_{\bolk}$ satisfies eq.(\ref{Isingtheta1}), the off-diagonal part of quadratic Hamiltonian is given by
\begin{equation}
\begin{split}
H_{\rm (I2)off}&=\sum_\bolk \Bigl\{S(-\epsilon_0({\bf k})\sinh 2\theta_{\bf k}+t_0({\bf k})\cosh 2\theta_{\bf k})\\
+\frac{J_1^z}{2}&(\frac{1}{N}\sum_\bolq C_1(\bolq)\sinh\theta_{\bolq})C_1(\bolk)\Bigr\}
(\alpha_{\bf k}\beta_{-\bf k}+\alpha_{\bf k}^\dagger\beta_{-\bf k}^\dagger)
\end{split}
\end{equation}
Now, approximately $\sinh \theta_{\bolk}=t_0(\bolk)/2\epsilon_0(\bolk)+\Delta_{\bolk}$, where $\Delta_{\bolk}$ is $O(J_1^{z-1})$. Then, the leading order of $\Delta_\bolk$ is obtained and, as a result, $\theta_{\bolk}$ is given by
\begin{subequations}
\begin{align}
&\sinh \theta_{\bolk}^{\rm (I2)}=A_{\rm I} C_1(\bolk)\ ,\\
&\cosh \theta_{\bolk}^{\rm (I2)}=1+\frac{A_{\rm I}^2 }{2}C_1(\bolk)^2\ ,
\end{align}
\end{subequations}
where
\begin{equation}
A_{\rm I}=\frac{2J_1^\perp S}{4S(4J_1^z-3J_2^z)-J_1^z}\ .\label{AI}
\end{equation}
Now, the quadratic part of the Hamiltonian reads
\begin{equation}
H_{\rm (I2)0}=\sum_{\bolk}(\epsilon_{\rm I2}(\bolk)-h)\alpha^\dagger_{\bolk}\alpha_\bolk
+(\epsilon_{\rm I2}(\bolk)+h)\beta^\dagger_\bolk\beta_\bolk \ ,
\end{equation}
where
\begin{equation}
\begin{split}
\epsilon_{\rm I2}(\bolk)&=8SJ_1^z-6SJ_2^z-16J_1^zA_{\rm I}^2+12J_2^zA_I^2\\
&+(2SJ_2^\perp +2J_2^z A_{\rm I}^2 )C_2(\bolk)\\
+&(2(8SJ_1^z-6SJ_2^z -J_1^{z})A_{\rm I}^2-4SJ_1^\perp A_{\rm I})C_1(\bolk)^2\ .
\label{I2disp}
\end{split}
\end{equation}
Then, the minimum of dispersion is obtained at $\bolQ_{1}=(0,0,0)$ or $\bolQ_{2}=(\pi,\pi,\pi)$ as in the first-order result. The chemical potentials and the effective masses for each phase are given by the same way as in Sec.~\ref{sec;S1} (see eq.(\ref{muS1}) and (\ref{masscl})). 

Let us evaluate the 2-body interaction $\Gamma$ between the condensed bosons. 
As in the first order case in $1/J_1^{z}$, we need to calculate the kernel in the ladder-diagram (see Fig.\ref{Fig;ladder}). When the gap of the $\alpha$ boson closes, that of the $\beta$ bosons is $O(J_1^z)$. Then, the correlation of the $\beta$ boson remains short-ranged for low energies and $\VEV{T(\beta_{\bolk}\beta_\bolk^\dagger)(E\approx 0)}=O(1/J_1^{z})$. The effect of $\beta$ operator in the interaction term is at most $O(1/\sqrt{J_1^z})$. Therefore, the interaction part of the Hamiltonian which affect the kernel is obtained and is given in (\ref{HintIsing2}). 
Now, we shall evaluate the kernel. 
Before doing so, a remark is in order; 
at the second order of $1/J_1^{z}$ an infinite number of diagrams appear in the kernel. Hence, with the help of large $S$ expansion, we keep the term of the third-order of $S$ (up to 2-loop diagrams) and neglect the term of $O((J_1^{z})^{-1}S^{-3})$. In the selection of diagrams which contribute to the kernel, we do not view $J_2^z$ as a special contrary to the case of the quadratic Hamiltonian, for simplicity. As a result, in the second order of $S$, four diagrams and, in the third order of $S$, fourteen diagrams contribute to the kernel. The one-loop diagrams are given by $3\sim 6$ shown in Fig.\ref{Fig;spin2}, and the part of the 2-loop diagrams are shown in Fig.\ref{Fig;Ising2diagram}. When we evaluate the diagrams, we drop the term of $O((J_1^{z})^{-2})$ after the frequency of the propagator is integrated out. 
In this calculation, we maintain the terms $J_2^{z}$ which are readily obtained in the quadratic Hamiltonian,  even if the contribution of these term is $O(J_2^{z}(J_1^z)^{-2})$. Concretely, we use the gap of $\beta$ boson as $2S(8J_1^z-6J_2^z)$ and maintain $J_2^z$ of (\ref{AI}) and (\ref{I2disp}). We solve (\ref{laddereq}) by substituting the obtained kernel, and the interaction $\Gamma$ between condensed bosons is obtained. 
The detailed results are shown in Sec.~\ref{sec:lattice}.

\begin{figure}[H]
\begin{center}
\includegraphics[scale=0.4]{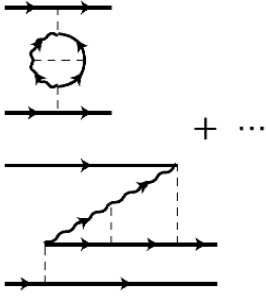}
\caption{The part of 2 loop diagrams which contributes to the kernel in the order in $S^{-2}(J_1^{z})^{-1}$. Straight lines (wavy lines) denote $\alpha\ (\beta)$ boson. Broken lines denote the momentum transfer at the interaction.
\label{Fig;Ising2diagram}
}
\end{center}
\end{figure}

\subsection{Possibility of bound-magnon BEC}
\label{2magBEC}
We briefly comment on the possibility of stable bound states. From the viewpoint of the Bethe-Salpeter equation, the two-particle Green's function contains the ladder diagram, and the divergence of $\Gamma$ implies the existence of stable bound states. In fact, this method has been successfully applied\cite{Chubukov,HTUandKT} to search for the stable bound-magnon state in the vicinity of the saturated ferromagnetic phase of 1D- and 3D frustrated
magnets. Hence, the above ladder-approximation may be also applied to
study the bound-magnon BEC around the solid. 
Now, we note that within the first order perturbation 
the effective interaction $\Gamma_{{\rm I1SS}_i}$ 
(eq.(\ref{eqn:GammaI1Qi})) diverges when 
\begin{equation}
1+0.258\frac{(J_2^z+|J_2^\perp|)}{S|J_2^\perp|}=0\ 
\; .\label{BMeq}
\end{equation}
In fact, we found that $\Gamma$ had a pole below the two-particle 
threshold when the left-hand side of eq.(\ref{BMeq}) is negative. 
Therefore, when the denominator of (\ref{eqn:GammaI1Qi}) 
is equal to 0 or negative, 
we may expect that, instead of the usual magnons, 
the bound magnon condenses 
at the center-of-mass momentum ${\bf K}=0$ 
provided that
the chemical potential is properly tuned. 
We found that the energy of the bound state 
at ${\bf K}=(\pi,\pi,\pi)$ is higher in energy than the one 
at ${\bf K}=0$ 
and will not affect the critical value of $\mu$ (or $h$) 
at which the bound-magnon BEC occurs. 
However, the extensive studies on hardcore bose Hubbard models by
quantum Monte-Calro 
simulations\cite{Batrouni-3,Batrouni-2,Dang,suzuki,Mila,dimerMC} 
have never indicated the existence of an SS accompanied by a bound-magnon BEC. 
Hence, it would be useful to reconsider this problem 
from the energetic point of view, even though rough. 

Of course, some other phases may compete with the bound magnon phase. 
In particular, the SF phase, which may appear from the solid phase 
via the first-order transition, would be an important candidate.
In the MFT, a solid-superfluid (S-SF) transition occurs at
\begin{equation}
\begin{split}
h_{\text{S-SF}}
&=2\sqrt{16J_1^{z2}-(3J_2^z+4|J_1^\perp|-3J_2^\perp)^2}\\
&= 8J_1^z-|O(J_1^{z-1})|\ .
\end{split}
\end{equation}
From (\ref{muI1}), on the other hand, 
one sees that the bound magnon BEC starts at $h_{b}\approx 8J_1^z-6(J_2^z+|J_2^\perp|)=8J_1^z+O((J_1^z)^0)$ when eq.(\ref{BMeq}) is satisfied. 
In the case of the attraction $(J_2^z+|J_2^\perp|)<0$, 
one sees that $h_{b} > h_{\text{S-SF}}$ and that 
a direct first-order S-SF transition occurs before the condensation 
of bound magnons. 
Hence, although an exotic SS phase brought about by the bound-magnon BEC
may be expected (note that the gap of a bound magnon closes 
earlier than that of a {\it single} magnon) 
in the vicinity of the solid phase, 
what we actually have is a phase separation.

Therefore, in order to see the bound-magnon BEC around the half-filled 
solid, it may be necessary that higher-order terms in the perturbation 
in $1/J_1^{z}$ shift the critical value $\mu$ by $O((J_1^z)^0)$. 
In this case, the approximation used in this section is beyond the scope of application to search the bound-magnon BEC. 
We do not go into more detailed discussion about the bound-magnon BEC in this paper.

\section{PHASE DIAGRAM}
\label{sec:lattice}
In Sec.~\ref{LargeS} and \ref{LargeJ}, 
we have described the two methods of calculating 
the minimum $\bolQ$ of the dispersion by which  
the spin structure of the SS phase has been determined. 
On top of it, the mass, the chemical potential, 
and the interaction $\Gamma$ which determines the stability of the SS 
have been computed. 
In this section, we show the detailed results on 
the phase diagram paying particular attention to 
the parameter set (\ref{LiuFisherSS}).

\subsection{Stability of a half-filled solid $-$a spinwave analysis}
\label{sec;resultS}
The solid phase is stable when the energy gap is finite.
Quantum fluctuations shift the energy gap and, in certain cases, the gap may close. In this subsection, we study the properties of the half-filled solid by the conventional spin-wave theory up to the second order in $S^{-1}$. Even for $S=1/2$, the approximation may work since the ground state is ordered. 

At the first order in $S^{-1}$, the energy gap closes even at $h=0$ when $|\tanh\theta_{\bolk=0}^{(1)}|=1$, and then the energy of the solid and the SF phase (or, a phase 
with magnetic long-range order in the $xy$-plane) is degenerate within the MFT (see (\ref{EIsing}), (\ref{ESF}) and (\ref{BogoC})). For $|\tanh\theta_{\bolk=0}^{(1)}|>1$, the SF phase is stabilized. 
In the second order in $S^{-1}$, the quantum fluctuation shift $\theta_{\bolk}^{(1)}$ to $\theta_\bolk^{(2)}$ and the boundary where the gap closes also changes. 
If the transition is a usual second-order one, the emergent phase may be SS. 
The first-order transition to the SF near the boundary may be also expected. However, in the case of 2D-square lattice, the quantum Monte-Calro simulations indicate that at the Mott-SF transition point, SU(2) symmetry dramatically restores\cite{Batrouni-2} as in the classical case. Even though there exists the difference of the dimensionality, we may not exclude the possibility that on the phase boundary SU(2) symmetry restores. 

To see the properties of the resultant phases more clearly, we carry out the Holstein-Primakoff transformation starting from the SF phase (the $xy$-ordered NAF phase in the spin language) for $J_1^\perp>0$ and calculate the magnon dispersion relation in the SF up to the second order in $S^{-1}$ at $h=0$. 

Let us briefly discuss some technical aspects of the calculation. There are two types of excitations: one is the gapless Goldston mode and the other is a massive (gapped) mode. 
To obtain these, we need to integrate out the functions of $\theta_{\bolk}$ 
as in (\ref{SpinT}). If we substitute the $\theta_{\bolk}$ obtained in the first order as in Sec.~\ref{sec:secondorder}, the gapless Goldston mode remains gapless. As is well known, the spin-wave expansion is well-behaved if the ground state is classically stable. Hence, we do not extend the calculation to the region where the corresponding phase is unstable in the MFT. 
Concerning the gapped mode, when the gap closes at the first order in $S^{-1}$, the solid and the SF are degenerate in energy within the MFT. 
Since the gap of this mode is affected by the quantum fluctuations, the phase boundary is shifted in the second order in $S^{-1}$. The resulting phase may be either the solid or the SS phase. 

As a result, a shift of the phase boundary is found in each phase, 
as is shown in Fig.\ref{Fig;spinwaveGround-1} and Fig.\ref{Fig;spinwaveGround-2}. 
As has been discussed above, since the spin-wave expansion is well-behaved in the case that the selected phase is the classical ground state, only on the classical phase boundary, we can compare the dispersion of each phase explicitly. 
At the first order in $S^{-1}$, both dispersions are gapless. 
At the second order in $S^{-1} $, when the dispersion of one phase (solid or SF) is ill-defined, that of the other phase obtains the finite gap. If the system has the global SU(2) (rotation) symmetry, the dispersion remains gapless. 
As a result, the shifted boundary forms the almost straight line which intersects that of the MFT at the parameter set where SU(2) symmetry exists.

\begin{figure}[H]
\begin{center}
\includegraphics[scale=0.43]{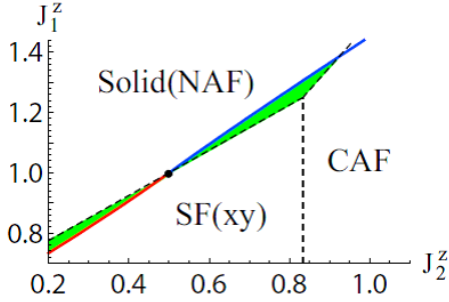}
\caption{(color online)
Phase boundary 
between the half-filled solid (or, the Ising-like NAF phase) and the SF phase 
(or, a phase with magnetic long-range order in the $xy$-plane) 
obtained by the large-$S$ expansion up to 
the order of $S^0$.  
Te values  $S=1/2,\ J_1^\perp=1,\ J_2^\perp=0.5$ and $h=0$
(half-filled) are used. 
The solid line (blue) for $J_2^z\geq 0.5$ denotes the boundary 
where the gap of the solid phase closes. 
The solid line (red) for $J_2^z\leq 0.5$ 
denotes the boundary where the gap of the massive mode of the SF phase closes.  
The broken line (black) denotes the classical boundary between NAF, CAF and SF phases. The dot represents the $J_1^z=1$ and $J_2^z=0.5$, where the system has $SU(2)$ symmetry and the boundaries intersect.
In the highlighted region (green), the dispersion has a non-zero imaginary part and is ill-defined. For $J_2^z\geq 0.5$, the emergent phase may be either the SF or the SS. For $J_2^z\leq 0.5$, the emergent phase may be either the solid or the SS. 
\label{Fig;spinwaveGround-1}
}
\end{center}
\end{figure}

\begin{figure}[H]
\begin{center}
\includegraphics[scale=0.43]{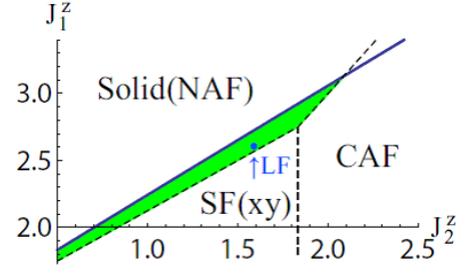}
\caption{(color online)
Phase boundary obtained by the large-$S$ expansion 
up to the order of $S^0$ between the half-filled solid and the SF phase for $S=1/2,\ J_1^\perp=-1,\ J_2^\perp=-0.5$ and $h=0$ (half-filled).
The solid line (blue) denotes the boundary where the gap of the solid phase closes. The broken line (black) denotes the classical boundary between NAF, CAF and SF phases. The dot labeled as LF (blue) represents the LF point (\ref{LiuFisherSS}), which is suggested for the fitting parameters\cite{Liu-Fisher} of $^4$He.
In the highlighted region (green), the spin-wave expansion is ill-defined and the emergent phase may be either the SF or the SS. The straight line and the broken line intersect at $J_1^z=1,\ J_2^z=-0.5$, where $SU(2)$ symmetry exists.
\label{Fig;spinwaveGround-2}
}
\end{center}
\end{figure}

Next, let us discuss the application to $^4$He as the QGM. 
As shown in Fig.\ref{Fig;spinwaveGround-2}, at the LF point, the solid phase is unstable even at $h=0$ and the resulting phase may be either the SF or the SS. To conjecture this phase, we plot the $\tanh\theta_{\bolk=0}$ of the massive modes as shown in Fig.\ref{Fig;tanh}. If $|\tanh\theta_{\bolk=0}|=1$, the energy gap closes. With the help of the fitting line, we see that on the parameters (3), the SF phase may be stabilized. Hence, on this parameter set, the QGM does not make a sense, and the fitting parameters for $^4$He must be reconsidered by taking into account the quantum fluctuation.

\begin{figure}[H]
\begin{center}
\includegraphics[scale=0.43]{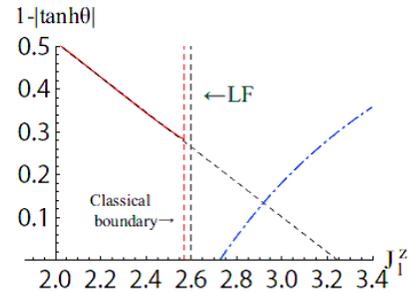}
\caption{(color online) 
$1-|\tanh\theta_{\bolk=0}^{(2)}|$ obtained of the order of $S^0$ for $S=1/2,\ J_2^z=1.59,\ J_1^\perp =\pm 1,\ J_2^\perp =-0.5$ 
plotted as a function of $J_1^z$. If $1-|\tanh\theta_{\bolk=0}^{(2)}|=0$, the gap closes. The solid line (red) is given by the massive mode on the SF. We obtained it for $J_1^\perp =1$, which transforms to $J_1^\perp =-1$ by the gauge transformation. The dashed line (blue) is obtained on the solid phase. The vertical line labeled as LF represents the LF point (\ref{LiuFisherSS}). On the classical boundary, the gap of the SF phase largely opens. If we introduce the fitting line (the non-labeled broken line), the gap of the SF phase seems to be open on the LF point and the stability of the SF is implied. 
Moreover, the gap seems to be maintained over the point where the gap of the solid phase closes. Hence, the solid-SF first-order transition is expected at $2.7 \lesssim J_1^z \lesssim 3.2$. 
\label{Fig;tanh}
}
\end{center}
\end{figure}

\subsection{stability of supersolid}
The fitting parameters for $^4$He shall shift from (\ref{LiuFisherSS}). Although the shift may be quantitatively large to see Fig.\ref{Fig;tanh}, that is still expected to be perturbative since the quantum fluctuation is treated as a perturbation. Then, since the effective interaction $\Gamma$ obtained within the first-order in $S^{-1}$ (or the MFT) on (\ref{LiuFisherSS}) is robust (see eq.(\ref{S1GammaSS1})), the perturbative shift of the fitting parameters shall not affect the stability of the SS within the MFT. Therefore, we study the quantum effect to the stability of the SS near the LF point (\ref{LiuFisherSS}) by the $\Gamma$ obtained in the second order perturbation in $S^{-1}$ and $1/J_1^{z}$. The repulsive nature of the effective interaction $\Gamma$$(>0)$ suggests the stability of the SS phase. 

Since $J_1^\perp/J_2^\perp$ is fixed at $1/2$ in the QGM, we plot $\Gamma$ as a function of $J_1^z$ and $J_2^z$, as is shown in Fig.\ref{Fig;GammaS2I2-2} and Fig.\ref{Fig;GammaS2I2}. On the phase boundary, the perturbation theory in $S^{-1}$ gives the divergence to $-\infty$ because of $\theta_{\bolk}$ for $\tanh\theta_{\bolk=0} \rightarrow 1$ and the used approximation is beyond control. Near (\ref{LiuFisherSS}), the perturbation theory in $1/J_1^{z}$ also has the problem of accuracy since the suppression of the expansion parameter $1/J^z_{1}$ may not be sufficient. However, both methods lead to the one identical conclusion. In the case of $S=1/2$, both predict that $\Gamma$ is considerably suppressed near (\ref{LiuFisherSS}), and, as a result, the second order term has the same magnitude as the relatively large first-order term. Hence, quantitatively, it may be understood that the $\Gamma$ of the MFT (that of the first order in $S^{-1}$) is not reliable near (\ref{LiuFisherSS}) and there exists the possibility that the quantum fluctuation breaks the stability of the SS. 
Therefore, even if the shift of the parameter set from (\ref{LiuFisherSS}) is perturbative, the stability of the SS phase of $^4$He remains to be a question. 

\begin{figure}[H]
\begin{center}
\includegraphics[scale=0.43]{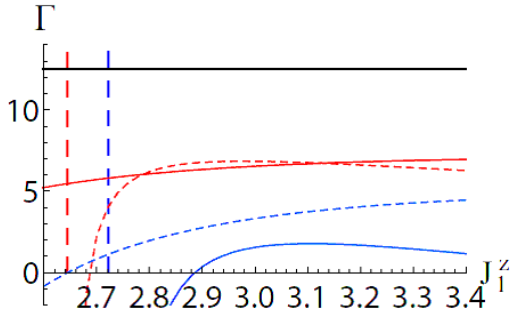}
\caption{(color online) The interaction $\Gamma$ 
for $J_2^z=1.59,\ J_1^\perp =-1,\ J_2^\perp =-0.5$. 
Solid lines are obtained in the second order in $S^{-1}$. The broken lines are in the second order in $(J_1^{z})^{-1}$. 
The curves are obtained respectively
 for $S=\infty$ (black), $1$ (red), $1/2$ (blue) beginning at the top. 
The left vertical line (red) is the phase boundary for $S=1$ 
where the gap closes in the second order in $S^{-1}$ at $h=0$. 
The right vertical line (blue) is for $S=1/2$.
Near the boundary and for large $J_1^z$, the difference becomes large.
For the large Ising-like anisotropy $J_1^z$, the evaluation of $\Gamma$
 in the second order in $S^{-1}$ becomes pathologic 
as discussed in the last part of Sec.~\ref{sec:secondorder}
\label{Fig;GammaS2I2-2}
}
\end{center}
\end{figure}

\begin{figure}[H]
\begin{center}
\includegraphics[scale=0.43]{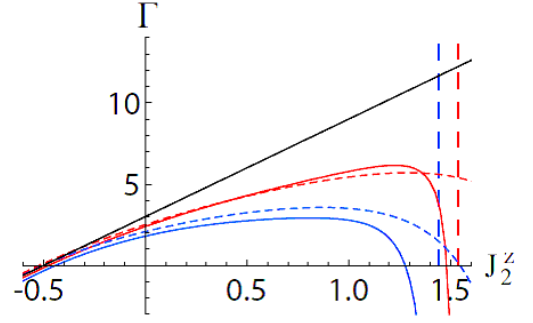}
\caption{(color online)  The interaction $\Gamma$ for $J_1^z=2.60,\ J_1^\perp =-1,\ J_2^\perp =-0.5$.
Solid lines are obtained in the second order in $S^{-1}$. The broken lines are in the second order in $(J_1^{z})^{-1}$. 
The curves are obtained respectively for $S=\infty$ (black), $1$ (red), $1/2$ (blue) beginning at the top. 
The right vertical line (red) is the phase boundary for $S=1$ where the gap closes in the second order in $S^{-1}$ at $h=0$. The left vertical line (blue) is for $S=1/2$. The shift of the boundary near $J_2^{z}=-0.5$ which determines the stability of the SS is extremely small and is within the error of $O(S^{-2})$ or $O((J_1^{z})^{-2})$. 
\label{Fig;GammaS2I2}
}
\end{center}
\end{figure}

Finally, we comment on the shift of the boundary which determines the stability of the SS given by the MFT. As is seen in Fig.\ref{Fig;GammaS2I2}, the quantum effect to $\Gamma$ is very little on this boundary and the shift is within the error of approximation for both approximation ($O(S^{-2})$ or $O((J_1^{z})^{-2})$). We found that this is also the case for the SS2 which mainly appears for $J_2^\perp>0$ unless the parameter set sits near the phase boundary of solid-SF at half filling. Therefore, if the energy of the solid phase is sufficiently less than the superfluid phase at half filling, the boundary determining the stability of the SS phase given by the MFT may not be affected by the quantum fluctuation.

\section{summary}
By using the spin-wave- ($1/S$) and the Ising expansion 
together with the dilute-Bose-gas technique, 
we studied the SS phase around the half-filled solid (NAF) phase. 
First, we introduced two kinds of magnon excitations for 
the two sublattices in the NAF phase. 
At a certain value of chemical potential 
(or, the external magnetic field), the gap of one of 
the Bogoliubov-transformed magnons closes;  
this magnon BEC keeps the two-sublattice NAF structure intact 
implying the SS phase. 
The spin configuration of the SS phase was 
 determined by the minimum 
of the energy spectrum over the solid ground state. 
The Bogoliubov-transformed magnons can be viewed as vacancies or 
interstitials introduced in solids. 
If the effective interaction $\Gamma$ among the condensed magnons 
are repulsive, we may expect, on physical grounds, a stable SS 
phase to appear. 
Therefore, the necessary condition for a second-order solid-SS 
transition is given by $\Gamma>0$; if this condition is met, 
the SS phase realizes for low condensate density. 

To evaluate the excitation spectrum in the solid phase 
and the effective interaction $\Gamma$ in a 
quantum-mechanical manner, 
we developed the perturbation theory in $S^{-1}$ and $(J_1^{z})^{-1}$ 
in Sec.~\ref{LargeS} and \ref{LargeJ}, respectively. 

\begin{table}
\caption{Types of phase transitions suggested by the several methods in this paper. 
`SW' and `Ising' represent the spinwave- and the Ising expansion discussed in Sec.~\ref{LargeS} and Sec.~\ref{LargeJ}, respectively. 
The interactions $\Gamma_{{\rm S1SSi}}$ ($i=1,2$) and 
$\Gamma_{{\rm I1SS}_i}$ are 
given in eqs.(\ref{S1GammaSS1}), (\ref{S1GammaSS2}) and 
(\ref{eqn:GammaI1Qi}), respectively. 
$\Gamma_{{\rm S2SSi}}$ and $\Gamma_{{\rm I2SS}_i}$ are shown in Figs.\ref{Fig;GammaS2I2-2},\ref{Fig;GammaS2I2}. 
If $\Gamma$ diverges, one should not take the value literally, 
since even in that case a second-order solid-`bound-magnon SS (BMSS)' transition may be expected. The detailed discussion on a BMSS is given in Sec.~\ref{2magBEC}. 
}
\label{table:phasetransition}
\begin{center}
\begin{ruledtabular}
\begin{tabular}{cccc}
\raisebox{0.5ex}[0pt]{method} & 
\raisebox{0.5ex}[0pt]{solid-SF} & 
\raisebox{0.5ex}[0pt]{solid-SS} & 
\raisebox{0.5ex}[0pt]{solid-BMSS}\\ 
\hline
SW 1st & 1st ($\Gamma_{{\rm S1SSi}}<0$) 
& 2nd ($\Gamma_{{\rm S1SSi}}>0$) & - \\ 
SW 2nd & 1st ($\Gamma_{{\rm S2SSi}}<0$) & 2nd ($\Gamma_{{\rm S2SSi}}>0$) & - \\ 
Ising 1st & 1st ($\Gamma_{{\rm I1SS}_i}<0$) & 2nd ($\Gamma_{{\rm I1SS}_i}>0$) & 2nd \\
Ising 2nd & 1st ($\Gamma_{{\rm I2SS}_i}<0$) & 2nd ($\Gamma_{{\rm I2SS}_i}>0$) & 2nd \\
MFT & 1st & 2nd & - \\ 
\end{tabular}

\end{ruledtabular}
\end{center}
\end{table}

The first-order calculation in $S^{-1}$ yielded  
the same results as in the MFT; three types of SS phases 
are found around the half-filled solid. 
At the second order in $S^{-1}$, we found a possibility that 
quantum fluctuations destabilize the NAF solid, which is expected 
to be stable from the MFT. 
Specifically, in the evaluation of $\Gamma$, 
the second-order correction in $S^{-1}$ becomes 
ill-behaved when the Ising-like anisotropy $J_1^z$ 
is large or when the energy of the solid is almost the same 
as that of the SF at half filling. 
In such cases, the MFT 
(or, equivalently, the first-order perturbation in $S^{-1}$) 
may not be reliable. 

In order to overcome this difficulty, 
we carried out another perturbation theory from 
the Ising limit (i.e. expansion in $1/J_1^{z}$). 
At the first order in $1/J_1^{z}$, 
we used the ladder approximation and saw that quantum 
fluctuations did suppress $\Gamma$, 
while we obtained the same result as the MFT one 
as far as the stability of the SS phase is concerned. 

For negative $\Gamma$ (i.e. attraction), we found a possibility 
of a novel phase characterized by the bound-magnon condensate. 
However, this phase may be replaced by the SF 
for the parameters considered in the text. 
In carrying out the second order calculation 
in $1/J_1^{z}$, 
we used the ladder approximation 
with only diagrams up to 2-loop (i.e. up to the third order in
$S^{-1}$) kept in the kernel. 
The effect of quantum fluctuations depends crucially on the energies of the 
solid- and the SF phase, as in the large-$S$ expansion.

When the energy of the solid is sufficiently smaller 
than that of the SF phase at half filling, 
the second order term had little effect on $\Gamma$ 
in the vicinity of the MFT-boundary 
(see Fig.\ref{Fig;GammaS2I2}). 
In other words, under the above condition, we may conclude that 
quantum fluctuations only have minor effects 
on the stability of the SS phase.

On the other hand, when the energy of the solid phase is comparable 
to that of the SF phase, 
there exists a possibility 
that quantum fluctuations completely wash out the SS phase 
obtained in the MFT. Actually, in the vicinity of the LF point, 
where frustration due to the competition among NAF, CAF and SF 
is strong, the second-order Ising-like expansion also 
concluded divergingly large negative values 
of $\Gamma$ (see Fig. \ref{Fig;GammaS2I2-2} 
and Fig. \ref{Fig;GammaS2I2}). 

In Sec.~\ref{sec:lattice}, we studied the effect of quantum fluctuations 
on the ground state at the LF point. 
At the second order in $S^{-1}$, the ground state may be 
given not by the SS but by the SF even at $h=0$. 
The failure of the Liu-Fisher values to describe $^4$He 
suggests that the optimal parameters, which should be 
obtained by fully quantum treatment, 
may differ from the Liu-Fisher ones.  
We expect that the deviation from 
the LF point is small and that it can be handled in a perturbative fashion. 
On the basis of this expectation, 
we studied the stability of the SS in the vicinity of the LF point. 
Since the energies of SS and SF are comparable in this region, 
the MFT may not be reliable. 
Even if the shift of the fitting parameters from the LF point 
is small and the MFT guarantees the stability of the SS, 
there remains a possibility that quantum fluctuations 
destabilize the SS. 
To investigate this possibility more closely, 
we shall need such a sophisticated treatment that 
the renormalization of the effective interaction $\Gamma$ 
due to higher order terms is appropriately taken into account. 

{\it Note added} -  
After the completion of our work, we became aware of a series of papers by Stoffel and Gul\'acsi who studied the same model\cite{Stoffel} as ours by the Green's function theory with the random-phase approximation. 
They reached a different conclusion that the critical external field at the solid-SS transition is little affected by quantum fluctuations at the LF point. 
We suspect that the discrepancy might be attributed to 
the difference in the approximation schemes; we believe that our 
approximation is well controlled by the two small parameters.

\begin{acknowledgements}
We thank T.~Momoi, N.~Shannon and D.~Yamamoto 
for useful discussions and helpful correspondences. 
H.T.U is grateful to the hospitality of Condensed Matter Theory 
Laboratory and the financial support from 
the Junior Research Associate program at RIKEN. 
The author (K.T.) was supported by Grant-in-Aid 
for Scientific Research (C) 20540375 and that on 
Priority Areas ``Novel States of Matter Induced by Frustration'' 
(No.19052003) from MEXT, Japan. 
This work was also supported by the Grant-in-Aid for the Global 
COE Program 
``The Next Generation of Physics, Spun from Universality 
and Emergence" from MEXT of Japan.
\end{acknowledgements}

\appendix
\section{Supersolid phase emerging from two types of bosons}
\label{AP;SS3}
In Sec.~\ref{sec;S1}, we have discussed the two SS phases, which
are described by a single Boson condensate, and classified them by $\Lambda$ (\ref{Lambda}). 
As has been mentioned there, however, in the case of $\Lambda=0$, the spin wave dispersion takes its minima at both ${\bf {\bf Q}}_1=(0,0,0)$ and ${\bf {\bf Q}}_2=(\pi,\pi,\pi)$, and there exists a possibility that both kinds of bosons condense {\it simultaneously}. 
In this appendix, we discuss this possibility within the first-order perturbation in $S^{-1}$. We shall see that a new type of SS phase (SS3) appears for a certain parameter region; it has a 4-sublattice structure and may continue to the quarter-filled solid.

As in the case of magnon BEC just below the saturation field\cite{Nikuni-Shiba-2,HTUandKT,VeilletteandChalker}, 
the ground-state energy density may be expanded in powers of the boson densities;
\begin{equation}
\begin{split}
\frac{E_{\text{eff}}}{N}&\approx \text{const}+\frac{1}{2}\Gamma_{\bolQ_1}\rho^2_{\bolQ_1}+\frac{1}{2}\Gamma_{\bolQ_2}\rho^2_{\bolQ_2}+\Gamma_2\rho_{\bolQ_1}\rho_{\bolQ_2}\\
+&\Gamma_3\rho_{\bolQ_1}\rho_{\bolQ_2}\cos 2(\varphi_{\bolQ_1}-\varphi_{\bolQ_2})-S\mu_0(\rho_{\bolQ_1}+\rho_{\bolQ_2})\ ,\label{sysE2}
\end{split}
\end{equation}
where
\begin{equation}
\begin{split}
\Gamma_2&=(V_{\alpha}(0;\bolQ_1,\bolQ_2)+V_{\alpha}(\bolQ_2-\bolQ_1;\bolQ_1,\bolQ_2)\\
&+V_{\alpha}(0;\bolQ_2,\bolQ_1)+V_{\alpha}(\bolQ_1-\bolQ_2;\bolQ_2,\bolQ_1))/2\ ,\\
\Gamma_3&=V_{\alpha}(\bolQ_2;\bolQ_1,\bolQ_1)\ (=V_{\alpha}(\bolQ_2;\bolQ_2,\bolQ_2))\ ,
\end{split}
\end{equation}
and $\VEV{\alpha_q}=\sqrt{N\rho_q}e^{i\varphi_q},\ \Gamma_{\bolQ_{j}}=\Gamma_{\rm S1SSj}$ for $j=1,2$ (eqs.(\ref{S1GammaSS1}) and (\ref{S1GammaSS2})).
Since the Hamiltonian (\ref{boseHorigin}) is not hermitian, it is not always true that $V_{\alpha}(\bolQ_2;\bolQ_1,\bolQ_1)= V_{\alpha}(\bolQ_2;\bolQ_2,\bolQ_2)$. However, these coincides with each other when $\Lambda=0$. The relative angle $(\varphi_{\bolQ_1}-\varphi_{\bolQ_2})$ takes 0 ($\pi/2$) when $\Gamma_3<0\ (>0)$. 

If $\Gamma_{\bolQ_i}<0$ or $\sqrt{\Gamma_{\bolQ_1}\Gamma_{\bolQ_2}}<-(\Gamma_2-|\Gamma_3|)$, a magnetization jump occurs. Otherwise, when $\text{Min}[\Gamma_{\bolQ_1},\Gamma_{\bolQ_2}]<\Gamma_2-|\Gamma_3|$, only one of the two species, which has smaller $\Gamma_{\bolQ_i}$ condenses and forms the spin structure (\ref{SSconfig}). 
If $\text{Min}[\Gamma_{\bolQ_1},\Gamma_{\bolQ_2}]>\Gamma_2-|\Gamma_3|$, (\ref{sysE2}) takes the minimum when
\begin{equation}
\begin{split}
\rho_{\bolQ_1}=\frac{\Gamma_{\bolQ_2}-(\Gamma_2-|\Gamma_3|)}{\Gamma_{\bolQ_1}\Gamma_{\bolQ_2}-(\Gamma_2-|\Gamma_3|)^2}S\mu_0\ ,\\
\rho_{\bolQ_2}=\frac{\Gamma_{\bolQ_1}-(\Gamma_2-|\Gamma_3|)}{\Gamma_{\bolQ_1}\Gamma_{\bolQ_2}-(\Gamma_2-|\Gamma_3|)^2}S\mu_0\ .
\end{split}
\end{equation}
Then, the spin configuration is given by,
\begin{subequations}
\begin{align}
\VEV{S_l^x}&=\sqrt{2S}(\sqrt{\rho_{\bolQ_1}}\cosh \theta^{(1)}_{{\bf Q}_1}\cos \varphi_{\bolQ_1}\nonumber\\
+&\sqrt{\rho_{\bolQ_2}}\cos ({\bf {\bf Q_2}}\cdot{\bf R}_l+\varphi_{\bolQ_2}))(1+\frac{f(\Delta S^{(1)})}{S})\ ,\nonumber\\
\VEV{S_l^y}&=\pm\sqrt{2S}(\sqrt{\rho_{\bolQ_1}}\cosh \theta^{(1)}_{{\bf Q}_1}\sin \varphi_{\bolQ_1}\nonumber\\
+&\sqrt{\rho_{\bolQ_2}}\sin ({\bf {\bf Q_2}}\cdot{\bf R}_l+\varphi_{\bolQ_2}))(1+\frac{f(\Delta S^{(1)})}{S})\ ,\text{for}\ \ l\in \text{A}\nonumber\\
\VEV{S^z_l}&=(S-\Delta S^{(1)})-(\rho_{\bolQ_1}\cosh^2\theta^{(1)}_{{\bf Q}_1}+\rho_{\bolQ_2}\nonumber\\
+&\cosh\theta^{(1)}_{\bolQ_1}\sqrt{\rho_{\bolQ_1}\rho_{\bolQ_2}}\cos ({\bf Q_2}\cdot{\bf R}_l+\varphi_{\bolQ_2}-\varphi_{\bolQ_1}))\ ,\\
\VEV{S_m^x}&=-\sqrt{2S\rho_{\bolQ_1}}\sinh \theta^{(1)}_{{\bf Q}_1}\cos \varphi_{\bolQ_1}(1+\frac{f(\Delta S^{(1)})}{S})\ ,\nonumber\\
\VEV{S_m^y}&=\mp\sqrt{2S\rho_{\bolQ_1}}\sinh \theta^{(1)}_{{\bf Q}_1}\sin \varphi_{\bolQ_1}(1+\frac{f(\Delta S^{(1)})}{S})\ ,\nonumber\\
\VEV{S^z_m}&=-(S-\Delta S^{(1)})+\rho_{\bolQ_1}\sinh^2\theta^{(1)}_{{\bf Q}_1}\ ,\\
&\ \ \ \ \ \text{for}\ \ m\in \text{B}\nonumber
\end{align}
\end{subequations}
where we use $\sinh\theta^{(1)}_{\bolQ_2}=0$ and $\Delta S$ and $f(\Delta S)$ is the same as in eq.(\ref{SSconfig}). By some numerical calculations, we found that this non-trivial SS phase with $(\varphi_{\bolQ_1}-\varphi_{\bolQ_2})=0$ (SS3) is stabilized for a broad region of the parameter-space, mainly for $J_2^z>0$. For example, if $J_1^z/|J_1^\perp|=3$ (and $\Lambda=0$), the SS3 exists for $0.2\lesssim  J_2^z/|J_1^\perp| \lesssim 2.0\ $.

For $\Lambda\approx 0$, $\Gamma_2$ and $\Gamma_3$ may have a influence on the magnetization process around the half-filled solid. For example, if $\Lambda >0$, the system energy is given by
\begin{equation}
\begin{split}
\frac{E_{\text{eff}}}{N}&\approx \frac{1}{2}\Gamma_{\bolQ_1}\rho^2_{\bolQ_1}+\frac{1}{2}\Gamma_{\bolQ_2}\rho^2_{\bolQ_2}+\Gamma_2\rho_{\bolQ_1}\rho_{\bolQ_2}\\
&+\Gamma_3\rho_{\bolQ_1}\rho_{\bolQ_2}\cos 2(\varphi_{\bolQ_1}-\varphi_{\bolQ_2})\\
&-S\mu_0\rho_{\bolQ_1}+(-S\mu_0 +\Delta_2)\rho_{\bolQ_2}\ ,\label{sysE2}
\end{split}
\end{equation}
where $\Delta_2=\epsilon_{\text{cl}}({\bf {\bf Q}}_2)-\epsilon_{\text{cl}}({\bf {\bf Q}}_1)\sim O(\Lambda)>0$ and $\Gamma$s obtained at $\Lambda=0$ may be used approximately. 
If the used parameters satisfy the condition of the stability of the SS3 discussed above, a phase transition from SS1 to SS3 occurs at:
\begin{equation}
S\mu_{0{\rm c1}}=\frac{\Gamma_{\bolQ_1}\Delta_2}{\Gamma_{\bolQ_1}-(\Gamma_2-|\Gamma_3|)}\ .
\end{equation}
Then, the densities of the condensed bosons are given by
\begin{subequations}
\begin{align}
\rho_{\bolQ_1}=\frac{(\Gamma_{\bolQ_2}-(\Gamma_2-|\Gamma_3|))S\mu_0+(\Gamma_2-|\Gamma_3|)\Delta_2}{\Gamma_{\bolQ_1}\Gamma_{\bolQ_2}-(\Gamma_2-|\Gamma_3|)^2}\ ,\\
\rho_{\bolQ_2}=\frac{(\Gamma_{\bolQ_1}-(\Gamma_2-|\Gamma_3|))S\mu_0-\Gamma_{\bolQ_1}\Delta_2}{\Gamma_{\bolQ_1}\Gamma_{\bolQ_2}-(\Gamma_2-|\Gamma_3|)^2}\ .
\end{align}
\label{tworho}
\end{subequations}
At $\mu_0=\mu_{0{\rm c1}}$, 
(\ref{tworho}) and (\ref{onerho}) give the same density $\rho_{\bolQ_{1,2}}$, and thus a second order phase transition is implied. 
If $\Lambda <0$, similarly, a second order phase transition from SS2 to SS3 occurs at:
\begin{equation}
S\mu_{0{\rm c}2}=\frac{\Gamma_{\bolQ_2}\Delta_1}{\Gamma_{\bolQ_2}-(\Gamma_2-|\Gamma_3|)}\ .
\end{equation}
where $\Delta_1=-\Delta_2$. 

\section{Some equations omitted in the text}
\label{sec:omitAP}
\subsection{Section \ref{sec:secondorder}}
The additional quadratic terms in eq.(\ref{Hqfull}) emerging from normal order of the Bogoliubov-transformed bosons are given by:
\begin{widetext}
\begin{equation}
\begin{split}
T_1({\bf k})
=&\epsilon_0 (\bolk)(\frac{1}{N}\sum_{\bolq}\sinh^2\theta_{q})-J_1^\perp ( \frac{1}{N}\sum_{\bolq}C_1({\bf \bolq})\sinh 2\theta_q)+(-\frac{2}{3}J_2^zC_2(\bolk)+2J_2^\perp )(\frac{1}{N}\sum_{\bolq}C_2({\bf q})\sinh^2\theta_{q})\ ,\\
T_2({\bf k})
=&t_0(\bolk)(\frac{1}{N}\sum_{\bolq}\sinh^2 \theta_q)-\frac{J_1^z}{4}C_1(k)(\frac{1}{N}\sum_{\bolq}C_1({\bf q})\sinh 2\theta_{q})\ ,\label{SpinT}
\end{split}
\end{equation}

\subsection{Section \ref{sec:Ising-2}}
The interaction part of Hamiltonian which contributes to the kernel of the order of $(J_1^{z})^{-1}$ is given by:
\begin{equation}
\begin{split}
H_{\rm (I2)int}=\frac{1}{N}\sum_{\bolq,\bolk_1,\bolk_2}&
\Bigl\{ (J_2^zC_2(\bolq)-J_2^\perp C_2(\bolk_2)-2J_1^z \sinh\theta_{\bolk_2-\bolq}\sinh\theta_{\bolk_2}C_1(\bolq)+J_1^\perp C_1(\bolk_2)\sinh\theta_{\bolk_2} )\alpha_{\bolk_1+\bolq}^\dagger\alpha_{\bolk_2-\bolq}^\dagger\alpha_{\bolk_1}\alpha_{\bolk_2}\\
&-2J_1^z C_1(\bolq)\alpha_{\bolk_1+\bolq}^\dagger\beta_{\bolk_2-\bolq}^\dagger\alpha_{\bolk_1}\beta_{\bolk_2}
+2J_1^zC_1(\bolq)\sinh\theta_{\bolk_2+\bolq}\alpha_{\bolk_1+\bolq}^\dagger\alpha_{\bolk_2+\bolq}\alpha_{\bolk_1}\beta_{-\bolk_2}\\
&+(2J_1^zC_1(\bolq)\sinh\theta_{\bolk_2-\bolq}-J_1^\perp C_1(\bolk_2))\alpha_{\bolk_1+\bolq}^\dagger\alpha_{\bolk_2-\bolq}^\dagger\alpha_{\bolk_1}\beta_{-\bolk_2}^\dagger\Bigr\}
\label{HintIsing2}
\end{split}
\end{equation}
\end{widetext}


\end{document}